\documentclass{SciPost}

\usepackage{bm}
\usepackage{dsfont}
\usepackage[framemethod=default]{mdframed}
\usepackage{showexpl}
\usepackage{booktabs}
\usepackage{soul}
\usepackage{alltt}
\usepackage{hyperref} 
\usepackage{algorithm}
\usepackage{algpseudocode}
\usepackage{cite}

\def\Tr{\mathop{\mathrm{Tr}}}
\def\Trf{\mathop{\mathrm{Tr}_{\mathrm{F}}}}
\definecolor{light-gray}{gray}{0.95}
\definecolor{dark-gray}{gray}{0.9}
\lstdefinestyle{fortran}{
  language=Fortran,
  basicstyle=\ttfamily,
  keywordstyle=\color{red},
  commentstyle=\color{blue},
  morecomment=[l]{!\ }
  breakatwhitespace=false,
  keepspaces=true,
  showstringspaces=false,
  columns=flexible,
  backgroundcolor=\color{light-gray},
  frame=single
}
\lstdefinestyle{bash}{
  language=bash,
  basicstyle=\ttfamily,
  keywordstyle=\color{red},
  commentstyle=\color{blue},
  morecomment=[l]{\#\ }
  breakatwhitespace=false,
  keepspaces=true,
  showstringspaces=false,
  columns=flexible
}
\DeclareRobustCommand{\hlgray}[1]{{\sethlcolor{dark-gray}\hl{#1}}}
\algnewcommand{\ThreeIndentedLeftComment}[1]{\Statex \(\hspace\algorithmicindent\hspace\algorithmicindent\hspace\algorithmicindent\triangleright\) \textit{#1}}
\algnewcommand{\FourIndentedLeftComment}[1]{\Statex \(\hspace\algorithmicindent\hspace\algorithmicindent\hspace\algorithmicindent\hspace\algorithmicindent\triangleright\) \textit{#1}}

\begin{document}

\begin{center}{\Large \textbf{
The \emph{ALF} (\emph{A}lgorithms for \emph{L}attice \emph{F}ermions) project release 1.0\\
Documentation for the  auxiliary field quantum Monte Carlo  code
}}\end{center}

\begin{center}
Martin Bercx,  Florian Goth,  Johannes S. Hofmann, Fakher F. Assaad 
\end{center}

\begin{center}
Institut f\"ur Theoretische Physik und Astrophysik, Universit\"at W\"urzburg, 97074 W\"urzburg, Germany
\\
alf@physik.uni-wuerzburg.de
\end{center}

\begin{center}
\today
\end{center}


\section*{Abstract}
{\bf 
The Algorithms for Lattice Fermions package provides a general code for the finite temperature auxiliary field quantum Monte Carlo algorithm.       The code  is engineered to  be able to simulate any model that can be written in terms of  sums of single-body operators, of squares of single-body operators and single-body operators coupled to an Ising field with  given dynamics. We  provide predefined types that allow  the user to specify the model, the  Bravais lattice  as well as equal time and time displaced observables.     The code supports an MPI implementation.   Examples such as the Hubbard model on the honeycomb lattice and  the Hubbard model  on the square lattice coupled to a transverse Ising field are  provided and discussed in the documentation.  We furthermore discuss  how to use the package  to implement  the Kondo lattice model and  the $SU(N)$-Hubbard-Heisenberg model.   One can download the code from our Git instance at  \path{https://alf.physik.uni-wuerzburg.de} and sign in to  file issues. 
}

\vfill
Copyright \textcopyright ~2016, 2017 The \textit{ALF} Project.\\
This is the ALF Project Documentation by the ALF contributors.
It is licensed under a Creative Commons Attribution-ShareAlike 4.0 International License.
You are free to share and benefit from this documentation as long as this license is preserved
and proper attribution to the authors is given. For details see the ALF project
homepage \url{alf.physik.uni-wuerzburg.de}. Contact address: \url{alf@physik.uni-wuerzburg.de} 
\clearpage

\noindent\rule{\textwidth}{1pt}
\tableofcontents\thispagestyle{fancy}
\noindent\rule{\textwidth}{1pt}
\clearpage

%
\section{Introduction}\label{sec:intro}
%
\subsection{Motivation}
%
The auxiliary field quantum Monte Carlo (QMC) approach is the algorithm of choice to simulate  thermodynamic properties of a variety of correlated electron systems in the solid state and beyond \cite{Blankenbecler81,White89,Sugiyama86,Sorella89, Duane87, Kennedy06, Assaad08_rev, gubernatis_kawashima_werner_2016}.  
Apart from the physics of the  canonical Hubbard model 
\cite{Scalapino07,LeBlanc15},   the topics one can investigate in detail include correlation effects in the bulk and on surfaces of topological insulators \cite{Hohenadler10,Zheng11}, quantum phase transitions between  Dirac fermions  and insulators \cite{Assaad13,Toldin14,Otsuka16,Chandrasekharan13,Chandrasekharan15},  
deconfined quantum critical points \cite{Li15a,Assaad16}, topologically ordered phases \cite{Assaad16}, heavy fermion systems \cite{Assaad99a,Capponi00}, nematic \cite{Schattner15} and magnetic  \cite{Xu16b} quantum phase transitions in metals, antiferromagnetism in metals \cite{Berg12},    superconductivity in spin-orbit split bands \cite{Tang14_1}, $SU(N)$ symmetric models \cite{Assaad04,Lang13},  long-ranged Coulomb interactions in graphene systems \cite{Hohenadler14,Tang15},  cold atomic gasses  \cite{Rigol03},  low energy nuclear physics \cite{Lee09},  entanglement entropies and spectra \cite{Grover13,Broecker14,Assaad13a,Assaad15, Broecker16},  etc. 
This ever growing list of topics  is based on  algorithmic progress and on recent symmetry related insights  \cite{Wu04,Huffman14,Yao14a,Wei16,Li16} enabling one to  find  negative sign  problem  free formulations of a number of model systems  with very rich phase diagrams.    

Auxiliary field methods  can be formulated in very different ways.  The fields define  the  configuration space $\mathcal{C}$. They can stem from the Hubbard-Stratonovich (HS)  \cite{Hubbard59} transformation required to decouple the  many-body interacting term into a sum of non-interacting problems,  or they can correspond to  bosonic modes with predefined dynamics such as phonons or gauge fields. In all cases, the result is that  the grand-canonical partition function  takes the form, 
\begin{equation}
	 Z = \text{Tr}\left( e^{-\beta \hat{\mathcal{H}}}\right)   =   \sum_{\mathcal{C}} e^{-S(\mathcal{C}) },
\end{equation}
where  $S$  is the action of non-interacting fermions subject to a  space-time fluctuating auxiliary field.    
The high-dimensional  integration  over the fields is carried out stochastically.  In this formulation of many  body quantum systems, there is no reason for the action to be a real number.  Thereby $e^{-S(\mathcal{C})}$ cannot be interpreted as a weight. To circumvent this problem one can adopt   re-weighting schemes and sample $| e^{-S(\mathcal{C})}| $. This invariably leads to the so called negative sign problem with associated exponential computational scaling in system size and inverse temperature \cite{Loh90, Troyer05}.    The  sign problem is formulation dependent, and as mentioned above there has been tremendous progress at identifying an ever growing class of  negative sign problem free models covering a  rich domain  of collective emergent  phenomena.  
 For continuous fields, the stochastic integrations can  be carried out with Langevin  dynamics or hybrid methods \cite{Duane85}.   However, for many  problems one can get away with discrete fields \cite{Hirsch83}. In this case,  Monte Carlo importance sampling will often be put to use \cite{Sokal89}.  
We note that  due to  the non-locality of the fermion determinant, see below, cluster updates,  such as in the loop or stochastic series expansion algorithms
 for quantum spin systems  \cite{Evertz93,Sandvik99b,Sandvik02}, are hard to formulate for this class of problems.  The search for efficient updating schemes that enable to move quickly within the configuration space defines ongoing challenges. 

 Formulations do not differ  only  by the choice of the fields, continuous or discrete,  and the sampling strategy, but also by the  formulation of the action itself.
For a given field configuration, integrating out  fermionic degrees of freedom generically leads to a fermionic determinant of dimension $\beta N$ where $\beta $  corresponds to the inverse temperature and $N$ to the volume of the system.  Working  with this determinant leads to the  Hirsch-Fye approach \cite{HirschFye86}  and its time complexity which quantifies the computational effort is given by  $\mathcal{O}\left( \beta N \right)^3$. \footnote{Here we implicitly assume the absence of negative sign problem}  The Hirsch-Fye  algorithm is the method of choice for impurity problems, but has  generically been outperformed by a class of so-called continuous-time quantum Monte Carlo approaches  
\cite{Gull_rev,Assaad14_rev, Assaad07}.    One key point of continuous-time methods  is that they are action based  and thereby allow to handle retarded interactions obtained when integrating out fermion or boson baths.  In high dimensions and/or at low temperatures, the cubic scaling originating from the fermionic determinant is expensive. To circumvent this, the hybrid Monte-Carlo approach  \cite{Duane87,Scalettar86}  expresses the fermionic determinant in terms of a Gaussian integral thereby introducing a new variable in the Monte Carlo integration.    The resulting algorithm is the method of choice for lattice gauge theories in 3+1 dimensions   and has been used to provide ab-inito  estimates of light hadron masses starting from quantum chromo dynamics \cite{Durr08}.

The algorithm implemented in the ALF project  lies between the  above two  \textit{extremes}.  We will keep the fermionic determinant, but formulate  the problem so as to  work only with $N\times N$ matrices.    This 
Blankenbecler,  Scalapino, Sugar (BSS)  algorithm scales linearly in  imaginary time $\beta$, but remains cubic in the volume $N$.    Furthermore, the algorithm can be formulated either in a projective manner \cite{Sugiyama86,Sorella89},  adequate to obtain zero temperature properties in the  canonical ensemble,  or at finite temperatures in the  grand-canonical ensemble \cite{White89}.

The aim of the ALF project is to introduce a general formulation of the \textit{ finite } temperature  auxiliary field QMC method with discrete  fields so as to quickly be able to play with different model Hamiltonians  at  minimal programming cost. We have summarized  the essential aspects of the   auxiliary field QMC  approach  in this documentation,   and   refer  the reader  to  Refs.~\cite{Assaad08_rev,Assaad02} for  complete  reviews.    
We will show in  all details how to implement a variety of models, run the code, and produce  results for  equal time and time displaced correlation functions. 
The program code is written in Fortran according to the 2003 standard and is able to natively utilize MPI for massively parallel runs on todays supercomputing systems.

The ALF package is  not the first open source project  aimed at providing simulation tools for correlated quantum matter.  The most  notable package is  certainly the ALPS library \cite{ALPS_2.0}.  It  is actively maintained and features a whole set of algorithms for strongly correlated quantum lattice models including Monte Carlo, exact diagonalization, and density matrix renormalization group codes. It however does not include the auxiliary field  QMC algorithm   offered by the ALF package.
 Other projects  include
QUEST \cite{QUEST}, TRIQS \cite{TRIQS}, w2dynamics \cite{w2dyn} and  iQist \cite{iQIST}.
IQist, TRIQS and w2dynamics focus on approximate solutions via the CT-HYB \cite{Gull_rev} algorithm within the dynamical mean field approximation.  
The QUEST project implements the same algorithm as in the ALF project  but  is currently restricted to the  Hubbard model
and it does not allow to easily incorporate different Hamiltonians.

The ALF source code  is placed  under the GNU GPL license.  The project is currently hosted on servers of the university of W\"urzburg  where 
 we have set up a GitLab instance (\url{https://alf.physik.uni-wuerzburg.de}) aimed at encouraging  community outreach. 
 Each  potential user can sign in, receive space for his ALF related projects and share them with others. This site serves the GitLab issue tracker  as well as  a wiki
so that members can collect information they consider useful for the project.
We have set up an E-Mail address for reaching the core developers at alf@physik.uni-wuerzburg.de.

%
\subsection{Definition of the Hamiltonian}
%
The first and most fundamental part of the project  is to define a general Hamiltonian which  can  accommodate a large class of models. 
Our approach is to express the model as a sum of one-body terms, a sum of two-body terms each written as a perfect square of a one body term, as well as a one-body term coupled to an Ising field with  dynamics to be specified by the user. 
The form of the interaction in terms of sums of perfect squares allows us to use generic forms of  discrete  approximations to the  HS  transformation \cite{Motome97,Assaad97}.
Symmetry considerations  are  imperative to enhance the speed of the code.  
We therefore include a \textit{color} index  reflecting  an underlying  $SU(N)$ color symmetry as  well as a flavor index  reflecting  the fact that  after  the HS  transformation,  the  fermionic determinant is block diagonal in this index.

The class of solvable models includes  Hamiltonians $\hat{\mathcal{H}}$ that have the following general form:
\begin{eqnarray}
\hat{\mathcal{H}}&=&\hat{\mathcal{H}}_{T}+\hat{\mathcal{H}}_{V} +  \hat{\mathcal{H}}_{I} +   \hat{\mathcal{H}}_{0,I}\;,\mathrm{where}
\label{eqn:general_ham}\\
\hat{\mathcal{H}}_{T}
&=&
\sum\limits_{k=1}^{M_T}
\sum\limits_{\sigma=1}^{N_{\mathrm{col}}}
\sum\limits_{s=1}^{N_{\mathrm{fl}}}
\sum\limits_{x,y}^{N_{\mathrm{dim}}}
\hat{c}^{\dagger}_{x \sigma   s}T_{xy}^{(k s)} \hat{c}^{\phantom\dagger}_{y \sigma s}  \equiv  \sum\limits_{k=1}^{M_T} \hat{T}^{(k)}
\label{eqn:general_ham_t}\\
\hat{\mathcal{H}}_{V}
&=&
\sum\limits_{k=1}^{M_V}U_{k}
\left\{
\sum\limits_{\sigma=1}^{N_{\mathrm{col}}}
\sum\limits_{s=1}^{N_{\mathrm{fl}}}
\left[
\left(
\sum\limits_{x,y}^{N_{\mathrm{dim}}}
\hat{c}^{\dagger}_{x \sigma s}V_{xy}^{(k s)}\hat{c}^{\phantom\dagger}_{y \sigma s}
\right)
+\alpha_{k s} 
\right]
\right\}^{2}  \equiv   
\sum\limits_{k=1}^{M_V}U_{k}   \left(\hat{V}^{(k)} \right)^2
\label{eqn:general_ham_v}\\
\hat{\mathcal{H}}_{I}  & = &
\sum\limits_{k=1}^{M_I} \hat{Z}_{k} 
\left(
\sum\limits_{\sigma=1}^{N_{\mathrm{col}}}
\sum\limits_{s=1}^{N_{\mathrm{fl}}}
\sum\limits_{x,y}^{N_{\mathrm{dim}}}
\hat{c}^{\dagger}_{x \sigma s} I_{xy}^{(k s)}\hat{c}^{\phantom\dagger}_{y \sigma s}
\right) \equiv \sum\limits_{k=1}^{M_I} \hat{Z}_{k}    \hat{I}^{(k)} 
\;.\label{eqn:general_ham_i}
\end{eqnarray}
The indices and symbols have the following meaning:
\begin{itemize}
\item The number of fermion \textit{flavors} is set by $N_{\mathrm{fl}}$.  After the HS transformation, the action will be block diagonal in the flavor index. 
\item The number of fermion \textit{colors} is set by $N_{\mathrm{col}}$.    The Hamiltonian is invariant under  $SU(N_{\mathrm{col}})$  rotations.~\footnote{Note that  in the code $ N_{\mathrm{col}} \equiv \texttt{N\_{SUN}} $.} 

\item Both the color and the flavor index can describe the spin degree of freedom, the choice depending on the spin symmetry 
of the simulated model and the HS transformation. This point is illustrated in the examples, see Secs.~\ref{sec:walk1} and \ref{sec:walk1.1}.

\item $N_{\mathrm{dim}}$ is the total number of spacial vertices: $N_{\mathrm{dim}}=N_{\text{unit cell}} N_{\mathrm{orbital}}$, 
where $N_{\text{unit cell}}$ is the number of unit cells of the underlying Bravais lattice and $N_{\mathrm{orbital}}$ is the number of (spacial) orbitals per unit cell.
\item The indices $x$ and $y$ label lattice sites where $x,y=1,\cdots, N_{\mathrm{dim}}$. 
\item Therefore, the  matrices $\bm{T}^{(k s)}$, $\bm{V}^{(ks)}$  and $\bm{I}^{(ks)}$ are  of dimension $N_{\mathrm{dim}}\times N_{\mathrm{dim}}$.
\item The number of interaction terms  is labelled by $M_V$   and $M_I$.   $M_T> 1 $ would allow for a checkerboard decomposition.
\item $\hat{c}^{\dagger}_{y \sigma s} $ is a second quantized operator that creates an electron in a Wannier state centered around lattice site $y$, with color $\sigma$, and  flavor index $s$.  The operators satisfy the anti-commutation relations: 
\begin{equation}
	\left\{ \hat{c}^{\dagger}_{y \sigma s},    \hat{c}^{\phantom\dagger}_{y' \sigma' s'}  \right\}   =   \delta_{y,y'}  \delta_{s,s'} \delta_{\sigma,\sigma'},   
	\; \text{ and } \left\{ \hat{c}^{\phantom\dagger}_{y \sigma s},    \hat{c}^{\phantom\dagger}_{y' \sigma' s'}  \right\}   =0.
\end{equation}

\end{itemize}
The Ising part of the general Hamiltonian (\ref{eqn:general_ham}) is $\hat{\mathcal{H}}_{0,I}+ \hat{\mathcal{H}}_{I}$ and  has the following properties:
\begin{itemize}
\item $\hat{Z}_k$ is an Ising spin operator which corresponds to the Pauli matrix $\hat{\sigma}_{z}$. It couples to a general one-body term. 
\item  The dynamics of the Ising spins is given by $\hat{\mathcal{H}}_{0,I}$. This term is not specified here; 
it has to be specified by the user and becomes relevant when the Monte Carlo update probability is computed in the code (see Sec.~\ref{sec:walk2} for an example).
\end{itemize}
Note that the matrices  $\bm{T}^{(ks)}$,  $\bm{V}^{(ks)}$ and  $\bm{I}^{(ks)}$ explicitly depend on the flavor index $s$ but not on the color index $\sigma$. 
The color index $\sigma$ only appears in  the  second quantized operators such that the Hamiltonian is manifestly $SU(N_{\mathrm{col}})$    symmetric.  We also require
the matrices $\bm{T}^{(ks)}$,  $\bm{V}^{(ks)}$ and  $\bm{I}^{(ks)}$  to be  Hermitian.

As we will detail below, the definition of the above Hamiltonian allows  to tackle  several  non-trivial models and phenomena.  There are however a number of model Hamiltonians that cannot be simulated with ALF.  Since we have opted for discrete  fields,   the electron-phonon interaction is not included. Furthermore, continuous HS transformations, that turn out to be extremely useful to include long-range Coulomb interactions \cite{Brower12,Ulybyshev2013,Hohenadler14},   are not accessible  in the present form of the package. \footnote{Note however that  one can readily add  short ranged interactions  by including terms such as $ (\hat{n}_i + \hat{n}_j - 2 )^2$. } In many cases such as in $^3$He,  three -  or more  body interactions  should be included to capture  relevant   exchange mechanisms \cite{Roger83,JWerner14a}. These higher order processes are   not captured in the ALF since it is \textit{limited}  to two-body interactions.  The formulation of the Hamiltonian, has an  explicit global $U(1)$ symmetry  corresponding to  particle number conservation.  Hence using the ALF for  a given model implies the existence of a canonical transformation where  a particle number is conserved.    Imaginary time dependent Hamiltonians, required to compute Renyi entropies and entanglement spectra   \cite{Broecker14,Assaad15, Broecker16} are not yet in the scope of ALF.  Finally, one should also mention that auxiliary field QMC simulations are Hamiltonian based such that retarded interactions  are not included in the ALF.    For this set of problems, CT-INT type approaches are the method of choice \cite{Assaad07,Werner07,Assaad14_rev}.     The above {\it short comings }  partially define a set of  future directions that will be discussed in the concluding part of this documentation.
%
\subsection{Outline}
%
To use the code, a minimal understanding of the algorithm is necessary. 
In Sec.~\ref{sec:def}, we go very briefly through  the steps required  to formulate the many-body imaginary-time propagation in terms of a sum  over HS and Ising fields  of one-body  imaginary-time propagators.   
The user has to provide this one-body imaginary-time propagator for a given configuration of   HS and  Ising fields. 
We equally discuss the Monte Carlo updates, the strategies for numerical stabilization of the code, as well as the Monte Carlo sampling.

Section \ref{sec:imp} is devoted to the data structures that are needed to implement the model, as well as to the input and output file structure.   
The data structure includes  an \texttt{Operator} type to  optimally work with sparse Hermitian matrices, a \texttt{Lattice} type  to define one- and two-dimensional Bravais lattices, and   two   \texttt{Observable} types to handle  scalar observables  (e.g. total energy)   and   equal time or time displaced two- point correlation functions (e.g. spin-spin correlations).

The Monte Carlo run and the  data analysis  are separated: the QMC run  dumps the results of \textit{bins}  sequentially into files  which are then analyzed by  analysis programs. In Sec.~\ref{sec:analysis}, we provide a brief description of the analysis programs  for our  observable types.  The analysis programs allow for omitting a given number of initial bins in order to account for warmup times. Also, a rebinning analysis is included  to a posteriori take  account of long autocorrelation times.  Finally, Sec.~\ref{sec:running} provides all the necessary details  to compile and run the code.

In Sec.~\ref{sec:ex}, we  give explicit examples on how to use the code for  the  Hubbard model on square and honeycomb lattices,  for different choices of the Hubbard-Stratonovich transformation  (see Secs.~\ref{sec:walk1},~\ref{sec:walk1.1} and ~\ref{sec:walk1.2})  as well as for the Hubbard model on a square lattice coupled to a transverse Ising field (see Sec.~\ref{sec:walk2} ).   Our implementation is rather general such that  a variety of other models can be simulated. In Sec.~\ref{sec:misc}   we provide  some information on how to simulate the Kondo lattice model as well as the $SU(N)$ symmetric Hubbard-Heisenberg model. 

Finally, in Sec.~\ref{sec:con} we list a number of features that are considered for  future releases of the ALF program package.

%
\section{Auxiliary Field Quantum Monte Carlo}\label{sec:def}
%
\subsection{Formulation of the method}  
%
Our aim is to compute observables  for the general Hamiltonian  (\ref{eqn:general_ham}) in
 thermodynamic equilibrium as described by the grand-canonical ensemble.
We will show below  how the grand-canonical partition function is rewritten as 
\begin{equation}
Z = \Tr{\left(e^{-\beta \hat{\mathcal{H}} }\right)}
= \sum_{C} e^{-S(C) } + \mathcal{O}(\Delta\tau^{2})
\end{equation}
and define the space of configurations  $C$.
Note that the chemical potential term is already included in the definition of the one-body term ${\mathcal{\hat{H}}_{T}}$, see eq. \eqref{eqn:general_ham_t}, of the general Hamiltonian.  

The outline of this section is as follows. First, we derive the detailed form of the partition function and outline the computation of observables (Sec.~\ref{sec:part} - \ref{sec:reweight}). 
Next, we present the present update strategy, namely local updates (Sec.~\ref{sec:updating}). 
We equally discuss the measures we have implemented to make the code numerically stable (Sec.~\ref{sec:stable}). Finally, we discuss the autocorrelations and associated time scales during the 
Monte Carlo sampling process (Sec.~\ref{sec:sampling}). 

The essential ingredients of the auxiliary field quantum Monte Carlo implementation in the ALF package are the following:
\begin{itemize}
\item  We will discretize the imaginary time propagation: $\beta = \Delta \tau L_{\text{Trotter}} $. Generically this introduces a systematic Trotter error of $\mathcal{O}(\Delta \tau)^2$  \cite{Fye86}. 
We note that there has been considerable effort at getting rid of the Trotter systematic error and to 
formulate a genuine continuous-time BSS  algorithm \cite{Iazzi15}.   To date, efforts in this direction are based on a CT-AUX type formulation \cite{Rombouts99,Gull08} and   face  two issues. The first issue is that they are restricted to a class of models with Hubbard-type interactions
\begin{equation}
        \left(  \hat{n}_{i}- 1\right)^{2}  = \left(  \hat{n}_{i}- 1\right)^{4} ,
\end{equation}
such that  the  basic CT-AUX equation \cite{Rombouts98}
\begin{equation}
          1   + \frac{U}{K} \left(  \hat{n}_{i}- 1\right)^{2}    = \frac{1}{2}\sum_{s=\pm 1}   e^{ \alpha s \left(  \hat{n}_{i}- 1\right) }  \; \text{ with  }  \;  \frac{U}{K} = \cosh(\alpha) -1 \; \text{ and  }  \; K\in\mathbb{R}
\end{equation}
holds.
The second issue is that in the continuous-time  approach it is hard to formulate a  computationally efficient algorithm.  Given this situation it turns out that the multi-grid method \cite{Rost12,Rost13,Bluemer08}  is an efficient  scheme to   extrapolate to  small imaginary-time steps so as to  eliminate the Trotter systematic error if required.
\item  Having isolated the two-body term,  we will use  the   discrete HS transformation \cite{Motome97,Assaad97}:
\begin{equation}
\label{HS_squares}
        e^{\Delta \tau  \lambda  \hat{A}^2 } =
        \sum_{ l = \pm 1, \pm 2}  \gamma(l)
e^{ \sqrt{\Delta \tau \lambda }
       \eta(l)  \hat{A} }
                + {\cal O} (\Delta \tau ^4)\;,
\end{equation}
where the fields $\eta$ and $\gamma$ take the values:
\begin{eqnarray}
 \gamma(\pm 1)  = 1 + \sqrt{6}/3, \quad  \eta(\pm 1 ) = \pm \sqrt{2 \left(3 - \sqrt{6} \right)}\;,\\\nonumber
  \gamma(\pm 2) = 1 - \sqrt{6}/3, \quad  \eta(\pm 2 ) = \pm \sqrt{2 \left(3 + \sqrt{6} \right)}\;.
\nonumber
\end{eqnarray}
Since the Trotter error is already of order $(\Delta \tau ^2) $ per time slice, this transformation is next to exact. 
\item  We will work in  a basis for the Ising spins  where  $\hat{Z}_k$ is diagonal: $\hat{Z}_{k}|s_{k}\rangle = s_{k}|s_{k}\rangle$, where $s_{k}=\pm 1$.
\item From the above it follows that the  Monte Carlo configuration space $C$  
is given by the combined spaces of Ising spin configurations  and of HS discrete field configurations:
\begin{equation}
	C = \left\{   s_{i,\tau} ,  l_{j,\tau}  \text{ with }  i=1\cdots M_I,\;  j = 1\cdots M_V,\; \tau=1\cdots L_{\mathrm{Trotter}}  \right\}.
\end{equation}
Here, the Ising spins take the values  $s_{i,\tau} = \pm 1$ and  the HS fields take the values  $l_{j,\tau}  = \pm 2, \pm 1 $.
\end{itemize}
%
\subsubsection{The partition function}\label{sec:part}
%
With the above, the partition function of the model (\ref{eqn:general_ham}) can be written as follows.
\begin{eqnarray}\label{eqn:partition_1}
Z &=& \Tr{\left(e^{-\beta \hat{\mathcal{H}} }\right)}\nonumber\\
  &=&   \Tr{  \left[ e^{-\Delta \tau \hat{\mathcal{H}}_{0,I}}   
    \prod_{k=1}^{M_V}   e^{ - \Delta \tau  U_k \left(  \hat{V}^{(k)} \right)^2}   \prod_{k=1}^{M_I}   e^{  -\Delta \tau  \hat{\sigma}_{k}  \hat{I}^{(k)}} 
    \prod_{k=1}^{M_T}   e^{-\Delta \tau \hat{T}^{(k)}}  
   \right]^{L_{\text{Trotter}}}}  + \mathcal{O}(\Delta\tau^{2})\nonumber \\
   &=&
   \sum_{C} \left( \prod_{k=1}^{M_V} \prod_{\tau=1}^{L_{\mathrm{Trotter}}} \gamma_{k,\tau} \right) e^{-S_{0,I} \left( \left\{ s_{i,\tau} \right\}  \right) }\times \nonumber\\
   &\quad&
    \Trf{ \left\{  \prod_{\tau=1}^{L_{\mathrm{Trotter}}} \left[   
    \prod_{k=1}^{M_V}   e^{  \sqrt{ -\Delta \tau  U_k} \eta_{k,\tau} \hat{V}^{(k)} }   \prod_{k=1}^{M_I}   e^{  -\Delta \tau s_{k,\tau}  \hat{I}^{(k)}} 
    \prod_{k=1}^{M_T}   e^{-\Delta \tau \hat{T}^{(k)}}    \right]\right\} }+ \mathcal{O}(\Delta\tau^{2})\;.
\end{eqnarray}
In the above,  the trace $\mathrm{Tr} $  runs over the Ising spins as well as over the fermionic degrees of freedom, and $ \mathrm{Tr}_{\mathrm{F}}  $ only over the  fermionic Fock space. 
$S_{0,I} \left( \left\{ s_{i,\tau} \right\}  \right)  $ is the action  corresponding to the Ising Hamiltonian,  and is only dependent on the Ising spins so that  it can be pulled out of the fermionic trace.  We have adopted the short hand notation $\eta_{k,\tau}  = \eta(l_{k,\tau})$   and $\gamma_{k,\tau}  = \gamma(l_{k,\tau})$.
At this point,  and  since for a given configuration $C$  we are dealing with a free propagation, we can integrate out the fermions to obtain a determinant: 
\begin{eqnarray}
 &\quad&\Trf{ \left\{  \prod_{\tau=1}^{L_{\mathrm{Trotter}}} \left[   
    \prod_{k=1}^{M_V}   e^{  \sqrt{ - \Delta \tau  U_k} \eta_{k,\tau} \hat{V}^{(k)} }   \prod_{k=1}^{M_I}   e^{  -\Delta \tau s_{k,\tau}  \hat{I}^{(k)}}  
    \prod_{k=1}^{M_T}   e^{-\Delta \tau \hat{T}^{(k)}}   \right] \right\}} = \nonumber\\
&\quad& \quad\prod\limits_{s=1}^{N_{\mathrm{fl}}} \left[  e^{\sum\limits_{k=1}^{M_V} \sum\limits_{\tau = 1}^{L_{\mathrm{Trotter}}}\sqrt{-\Delta \tau U_k}  \alpha_{k,s} \eta_{k,\tau} }
   \right]^{N_{\mathrm{col}}}\times
\nonumber\\
&\quad&\quad   \prod\limits_{s=1}^{N_{\mathrm{fl}}} 
   \left[
    \det\left(  \mathds{1} + 
     \prod_{\tau=1}^{L_{\mathrm{Trotter}}}  
    \prod_{k=1}^{M_V}   e^{  \sqrt{ -\Delta \tau  U_k} \eta_{k,\tau} {\bm V}^{(ks)} }   \prod_{k=1}^{M_I}   e^{  -\Delta \tau s_{k,\tau}  {\bm I}^{(ks)}}   \prod_{k=1}^{M_T}   e^{-\Delta \tau \textbf{T}^{(ks)}}  
     \right) \right]^{N_{\mathrm{col}}}
\end{eqnarray}
where the matrices $\textbf{T}^{(ks)}$,  $\textbf{V}^{(ks)}$, and  $\textbf{I}^{(ks)}$ define the Hamiltonian [Eq.~(\ref{eqn:general_ham}) - (\ref{eqn:general_ham_i})].
All in all,   the partition function is given by:
\begin{eqnarray}\label{eqn:partition_2}
    Z  \hspace{-0.7em}&=&  \hspace{-0.7em}   \sum_{C}   e^{-S_{0,I} \left( \left\{ s_{i,\tau} \right\}  \right) }     \left( \prod_{k=1}^{M_V} \prod_{\tau=1}^{L_{\mathrm{Trotter}}} \gamma_{k,\tau} \right)
    e^{ N_{\mathrm{col}}\sum\limits_{s=1}^{N_{\mathrm{fl}}} \sum\limits_{k=1}^{M_V} \sum\limits_{\tau = 1}^{L_{\mathrm{Trotter}}}\sqrt{-\Delta \tau U_k}  \alpha_{k,s} \eta_{k,\tau} } 
  \times   \nonumber \\
  \hspace{-0.7em}&\quad&  \hspace{-0.7em}
      \prod_{s=1}^{N_{\mathrm{fl}}}\left[\det\left(  \mathds{1} + 
     \prod_{\tau=1}^{L_{\mathrm{Trotter}}}   
    \prod_{k=1}^{M_V}   e^{  \sqrt{ -\Delta \tau  U_k} \eta_{k,\tau} {\bm V}^{(ks)} }   \prod_{k=1}^{M_I}   e^{  -\Delta \tau s_{k,\tau}  {\bm I}^{(ks)}}  
      \prod_{k=1}^{M_T}   e^{-\Delta \tau {\bm T}^{(ks)}}  
     \right) \right]^{N_{\mathrm{col}}} \hspace{-0.7em}+ \mathcal{O}(\Delta\tau^{2}) \nonumber\\
    \hspace{-0.7em}&\equiv&  \hspace{-0.7em}   
     \sum_{C} e^{-S(C) } + \mathcal{O}(\Delta\tau^{2})\;.
\end{eqnarray}
In the above, one notices that the weight factorizes in  the flavor index. The color index raises the determinant to the power $N_{\mathrm{col}}$. 
This corresponds to  an explicit $SU(N_{\mathrm{col}})$ symmetry   for each  configuration. This symmetry is manifest in the fact that the single particle  Green functions are color independent, again for each given  configuration $C$.
%
\subsubsection{Observables}\label{Observables.General}
%
In the auxiliary field QMC approach, the single-particle Green function plays a crucial role.  It determines the Monte Carlo dynamics and is used to compute  observables:
\begin{equation}\label{eqn:obs}
\langle \hat{O}  \rangle  = \frac{ \text{Tr}   \left[ e^{- \beta \hat{H}}  \hat{O}   \right] }{ \text{Tr}   \left[ e^{- \beta \hat{H}}  \right] } =   \sum_{C}   P(C) 
   \langle \langle \hat{O}  \rangle \rangle_{(C)} , \text{   with   } 
  P(C)   = \frac{ e^{-S(C)}}{\sum_C e^{-S(C)}}\;.
\end{equation}
$\langle \langle \hat{O}  \rangle \rangle_{(C)} $ corresponds to the expectation value of  $\hat{O}$ for a given configuration $C$.
For a given configuration $C$  one can use Wick's theorem to compute $ \langle \langle \hat{O}  \rangle \rangle_{(C)}  $   from the knowledge of the single-particle Green function: 
\begin{equation}
       G( x,\sigma,s, \tau |    x',\sigma',s', \tau')   =       \langle \langle \mathcal{T} \hat{c}^{\phantom\dagger}_{x \sigma s} (\tau)  \hat{c}^{\dagger}_{x' \sigma' s'} (\tau') \rangle \rangle_{C}
\end{equation}
where $ \mathcal{T} $ corresponds to the imaginary-time ordering operator.   The  corresponding equal time quantity reads, 
\begin{equation}
       G( x,\sigma,s, \tau |    x',\sigma',s', \tau)   =       \langle \langle  \hat{c}^{\phantom\dagger}_{x \sigma s} (\tau)  \hat{c}^{\dagger}_{x' \sigma' s'} (\tau) \rangle \rangle_{C}.
\end{equation}
Since  for a given HS field translation invariance in imaginary-time is broken, the Green function has an explicit $\tau$ and $\tau'$ dependence.   On the other hand it is diagonal in the flavor index, and independent on the color index. The latter reflects the  explicit $SU(N)$   color symmetry present at the level of individual HS configurations.   As an example,  one can show that the equal time Green function at $\tau = 0$ reads \cite{Assaad08_rev}:
\begin{equation}\label{eqn:Green_eq}
G(x,\sigma,s,0| x',\sigma,s,0 )  =   \left(  \mathds{1}  +  \prod_{\tau = 1}^{L_{\text{Trotter}}}  \bm{B}_{\tau}^{(s)}   \right)^{-1}_{x,x'}
\end{equation}
with
\begin{equation}
	\bm{B}_{\tau}^{(s)} =  \prod_{k=1}^{M_T}   e^{-\Delta \tau {\bm T}^{(ks)}}  
    \prod_{k=1}^{M_V}   e^{  \sqrt{ -\Delta \tau  U_k} \eta_{k,\tau} {\bm V}^{(ks)} }   \prod_{k=1}^{M_I}   e^{  -\Delta \tau s_{k,\tau}  {\bm I}^{(ks)}}.
\end{equation}

To compute equal time as well as time displaced observables,  one can make use of Wick's theorem. A convenient formulation of this theorem for  QMC simulations reads: 
\begin{eqnarray}
& & \langle \langle 	\mathcal{T}   c^{\dagger}_{\underline x_{1}}(\tau_{1}) c^{\phantom\dagger}_{{\underline x}'_{1}}(\tau'_{1})  
\cdots c^{\dagger}_{\underline x_{n}}(\tau_{n}) c^{\phantom\dagger}_{{\underline x}'_{n}}(\tau'_{n}) 
\rangle \rangle_{C} =
\nonumber \\ & & \det  
\begin{bmatrix}
   \langle \langle   \mathcal{T}   c^{\dagger}_{\underline x_{1}}(\tau_{1}) c^{\phantom\dagger}_{{\underline x}'_{1}}(\tau'_{1})  \rangle \rangle_{C} & 
    \langle \langle  \mathcal{T}   c^{\dagger}_{\underline x_{1}}(\tau_{1}) c^{\phantom\dagger}_{{\underline x}'_{2}}(\tau'_{2})  \rangle \rangle_{C}  & \dots   &   
    \langle \langle   \mathcal{T}   c^{\dagger}_{\underline x_{1}}(\tau_{1}) c^{\phantom\dagger}_{{\underline x}'_{n}}(\tau'_{n})  \rangle \rangle_{C}  \\
    \langle \langle   \mathcal{T}   c^{\dagger}_{\underline x_{2}}(\tau_{2}) c^{\phantom\dagger}_{{\underline x}'_{1}}(\tau'_{1})  \rangle \rangle_{C}  & 
      \langle \langle   \mathcal{T}   c^{\dagger}_{\underline x_{2}}(\tau_{2}) c^{\phantom\dagger}_{{\underline x}'_{2}}(\tau'_{2})  \rangle \rangle_{C}  & \dots  &
       \langle \langle   \mathcal{T}   c^{\dagger}_{\underline x_{2}}(\tau_{2}) c^{\phantom\dagger}_{{\underline x}'_{n}}(\tau'_{n})  \rangle \rangle_{C}   \\
    \vdots & \vdots &  \ddots & \vdots \\
    \langle \langle   \mathcal{T}   c^{\dagger}_{\underline x_{n}}(\tau_{n}) c^{\phantom\dagger}_{{\underline x}'_{1}}(\tau'_{1})  \rangle \rangle_{C}   & 
     \langle \langle   \mathcal{T}   c^{\dagger}_{\underline x_{n}}(\tau_{n}) c^{\phantom\dagger}_{{\underline x}'_{2}}(\tau'_{2})  \rangle \rangle_{C}   & \dots  & 
     \langle \langle   \mathcal{T}   c^{\dagger}_{\underline x_{n}}(\tau_{n}) c^{\phantom\dagger}_{{\underline x}'_{n}}(\tau'_{n})  \rangle \rangle_{C}
 \end{bmatrix}.
\end{eqnarray}
Here, we have defined the super-index $\underline{ x} = \left\{   x,\sigma,s \right\}$.
In the subroutines   \path{Obser}  and \path{ObserT} of  the module \path{Hamiltonian_Examples.f90} (see Sec.~\ref{sec:obs})   the user is provided with the equal time and time displaced correlation function. Using the  above  formulation  of  Wick's theorem, arbitrary  correlation functions can be computed. We note however, that the program is limited to the calculation of observables that contain only two different imaginary times.  
%
\subsubsection{Reweighting and the sign problem}\label{sec:reweight}
%
In general, the action  $S(C) $ will be complex, thereby inhibiting a direct Monte Carlo sampling of $P(C)$.   This leads to the infamous sign problem.     The sign problem is formulation dependent and as noted above, much progress has been made at understanding the class of models that  can be formulated without encountering this problem 
\cite{Wu04,Huffman14,Yao14a,Wei16}.  When the average sign is not too small, we can nevertheless  compute observables within a reweighting scheme.   Here we adopt the following scheme. First  note  that the partition function is real such that: 
\begin{equation}
	Z =   \sum_{C}  e^{-S(C)}    =  \sum_{C}  \overline{e^{-S(C)}} = \sum_{C}  \Re \left[e^{-S(C)} \right]. 
\end{equation}
Thereby\footnote{The attentive reader will have noticed that   for arbitrary Trotter decompositions,  the  imaginary time propagator is not necessarily Hermitian. Thereby, the above equation is correct only up to corrections stemming from the  controlled Trotter systematic error. }
and with the definition
\begin{equation}
\label{Sign.eq}
	 \text{ sign }(C)   =  \frac{   \Re \left[e^{-S(C)} \right]  } {\left| \Re \left[e^{-S(C)} \right]  \right|  }\;,
\end{equation}
the computation of the observable [Eq.~(\ref{eqn:obs})] is re-expressed as follows:
\begin{eqnarray}\label{eqn:obs_rw}
\langle \hat{O}  \rangle  &=&  \frac{\sum_{C}  e^{-S(C)} \langle \langle \hat{O}  \rangle \rangle_{(C)} }{\sum_{C}  e^{-S(C)}}       \nonumber \\ 
                          &=&  \frac{\sum_{C}   \Re \left[e^{-S(C)} \right]    \frac{e^{-S(C)}} {\Re \left[e^{-S(C)} \right]}  \langle \langle \hat{O}  \rangle \rangle_{(C)} }{\sum_{C}   \Re \left[e^{-S(C)} \right]}    \nonumber \\ 
          &=&
   \frac{
     \left\{
      \sum_{C}  \left| \Re \left[e^{-S(C)} \right]  \right|   \text{ sign }(C)   \frac{e^{-S(C)}} {\Re \left[e^{-S(C)} \right]}  \langle \langle \hat{O}  \rangle \rangle_{(C)}  \right\}/
            \sum_{C}  \left| \Re \left[ e^{-S(C)} \right] \right|  
          }  
          { 
          \left\{ \sum_{C}  \left|  \Re \left[ e^{-S(C)} \right]   \right|   \text{ sign }(C) \right\}/
            \sum_{C}   \left| \Re \left[ e^{-S(C)} \right] \right|  
          } \nonumber\\
          &=&
  	 \frac{  \left\langle  \text{ sign }   \frac{e^{-S}} {\Re \left[e^{-S} \right]}  \langle \langle \hat{O}  \rangle \rangle  \right\rangle_{\overline{P}} } { \langle \text{sign}   \rangle_{\overline{P}}}  \;.      
\end{eqnarray} 
The average sign is 
\begin{equation}\label{eqn:sign_rw}
	 \langle \text{sign} \rangle_{\overline{P}} =    \frac { \sum_{C}  \left|  \Re \left[ e^{-S(C)} \right]   \right|   \text{ sign }(C) }  {  \sum_{C}   \left| \Re \left[ e^{-S(C)} \right] \right|  } \;,
\end{equation}
and we have  $\langle \text{sign} \rangle_{\overline{P}} \in \mathbb{R}$ per definition.
The Monte Carlo simulation samples the probability distribution 
\begin{equation}  
	 \overline{P}(C) = \frac{ \left|  \Re \left[ e^{-S(C)} \right] \right| }{\sum_C \left|  \Re \left[ e^{-S(C)} \right]  \right| }\;.
\end{equation}
such that the nominator and denominator of  Eq.~(\ref{eqn:obs_rw})  can be computed.   

The negative sign problem is an issue since the average sign is a ratio of two partition functions such that one can argue that 
\begin{equation}
 \langle \text{sign} \rangle_{\overline{P}}   \propto e^{-  \Delta N \beta}.
\end{equation}
$\Delta $ is intensive positive quantity and $N \beta$ denotes the  Euclidean volume.    In a Monte Carlo simulation, the  error scales as $ 1/\sqrt{T_{\text{CPU}}} $   where $T_{\text{CPU}}$ corresponds to the computational  time.  Since the error on the  average sign has to be much smaller than the average sign itself,   one sees that:
\begin{equation}
	T_{\text{CPU}}  \gg e^{2 \Delta N \beta}.
\end{equation}   
Two comments are in order. First, the presence of a sign problem invariably leads to an exponential increase of CPU time as a function of the Euclidean volume. And second,  $\Delta$ is formulation dependent.  
For instance, at finite doping, the SU(2) invariant formulation of the Hubbard model presented in Sec.~\ref{sec:walk1}   has a much more severe sign problem than the formulation presented in Sec.~\ref{sec:walk1.1} where the HS field couples to the z-component of the magnetization.  Typically one can work with average signs  down to $ \langle \text{sign} \rangle_{\overline{P}} \simeq 0.1 $.

%
\subsection{Updating schemes}\label{sec:updating}
%
The program allows for different types of updating schemes.    Given a configuration $C$ we propose a new one, $C'$, with probability $T_0(C \rightarrow C')$  and accept it according to   the  Metropolis-Hastings   acceptance-rejection probability, 
\begin{equation}
	P(C \rightarrow C') =  \text{min}  \left( 1, \frac{T_0(C' \rightarrow C) W(C')}{T_0(C \rightarrow C') W(C)} \right),
\end{equation}
so as to guarantee the stationarity condition.  Here, $ W(C) = \left| \Re \left[ e^{-S(C)} \right] \right|   $.

\begin{table}[h]
   \begin{tabular}{@{} l l l @{}}\toprule
        Variable  &  Type                  &  Description   \\
         \\\midrule
       \texttt{Propose\_S0}   &    Logical       &  If true, proposes local moves according to the probability $e^{-S_{0,I}}$. 
         \\\bottomrule
   \end{tabular}
   \caption{Variable required to control the updating scheme. \label{table:Updating_schemes}}
\end{table}
%
\subsubsection{The default: sequential  single spin flips}
%
The default updating scheme is a  sequential single  spin flip algorithm.   Consider   the Ising spin $s_{i,\tau}$. We will flip it with probability one such that for  this local move  the  proposal matrix is symmetric.  If we are considering the Hubbard-Stratonovich field $l_{i,\tau}$  we will propose with probability $1/3$ one  of the other three  possible fields.   Again, for this local move, the proposal matrix is symmetric.  Hence in both cases we will accept or reject the move according to 
 \begin{equation}
 	P(C \rightarrow C') =  \text{min}  \left( 1, \frac{ W(C')}{W(C)} \right).
 \end{equation}
 It is worth noting that this type of sequential spin flip updating does not satisfy detailed balance but the more fundamental stationarity condition \cite{Sokal89}. 
%
\subsubsection{Sampling of $e^{-S_{0,I}}$}
%
Consider an Ising spin at space-time $i,\tau$ and the configuration $C$. Flipping this spin will generate the configuration $C'$ and we will propose the move according to 
  \begin{equation}
 T_0(C \rightarrow C')  =  \frac{e^{-S_{0,I}(C')}}{ e^{-S_{0,I}(C')} + e^{-S_{0,I}(C)} }   = 1 - \frac{1}{1 +  e^{-S_{0,I}(C')} /e^{-S_{0,I}(C)}}.
  \end{equation}
 Note that the function $\texttt{S0}$ in the  \texttt{Hamitonian\_example.f90}  module  computes precisely the ratio ${e^{-S_{0,I}(C')} /e^{-S_{0,I}(C)}}$ so that  $T_0(C \rightarrow C') $ does not require any further programming. 
 Thereby one will accept the proposed move with the probability: 
 \begin{equation}
 P(C \rightarrow C') =  \text{min}  \left( 1,  \frac{e^{-S_{0,I}(C)}   W(C')}{ e^{-S_{0,I}(C')} W(C)} \right).
 \end{equation}
 With Eq.~\ref{eqn:partition_2}  one sees that the bare action $S_{0,I}(C)$  determining the  dynamics of the Ising spin  in the absence of coupling to the fermions  does not enter the Metropolis acceptance-rejection step. This sampling scheme is used if the logical variable \texttt{Propose\_S0} is set to \texttt{true}.
%
\subsection{Stabilization - a peculiarity of the BSS algorithm}\label{sec:stable}
%
From \eqref{eqn:partition_2} it can be seen that for the calculation of the Monte Carlo weight
and for the observables a long product of matrix exponentials has to be formed.
On top of that we need to be able to extract the single-particle Green function  for a given flavor index at say time slice $\tau = 0$.  As  mentioned above in Eq.~(\ref{eqn:Green_eq}), this quantity is given by: 
\begin{equation}
\bm{G}= \left( \mathds{1} + \prod_{ \tau= 1}^{L_{\text{Trotter}}} \bm{B}_\tau \right)^{-1}.
\end{equation}
To boil this down to more familiar terms from linear algebra we remark that we can recast this problem as the task to find the solution of the linear system
\begin{equation}
(\mathds{1} + \prod_\tau \bm{B}_\tau) x = b.
\end{equation}
The $\bm{B}_\tau \in \mathbb{C}^{n\times n}$ depend on the lattice size as well as other physical parameters that can be chosen such that a matrix norm of $\bm{B}_\tau$ can be unbound in size.
From standard perturbation theory for linear systems it is known that the computed solution $\tilde{x}$ would 
contain a relative error of
\begin{equation}
\frac{|\tilde{x} - x|}{|x|} = \mathcal{O}\left(\epsilon \kappa_p\left(\mathds{1} + \prod_\tau \bm{B}_\tau\right)\right).
\end{equation}
Here $\epsilon$ denotes the machine precision, which is $2^{-53}$ for IEEE double precision numbers,
and $\kappa_p(\bm{M})$ is the condition number of the matrix $\bm{M}$ with respect to the matrix $p$-norm.
The important property that makes straight-forward inversion so badly suited  stems from the fact that $  \prod_ \tau \bm{B}_\tau $ contains exponentially large and small scales as can be seen in Eq.~\eqref{eqn:partition_2}.  Thereby, as a function of increasing inverse temperature, 
the condition number will grow exponentially so that the computed solution $\tilde{x}$
will often contain no correct digits at all.
To circumvent this, more sophisticated methods have to be employed. We will first of all assume that the multiplication of \texttt{NWrap} $\bm{B}$ matrices has an acceptable condition number.
Assuming for simplicity that $L_{\text{Trotter}}$ is an integer multiple of \texttt{NWrap}, we can write:
\begin{equation}\label{eqn:green_stable}
\bm{G} = \left( \mathds{1} + \prod\limits_{ i = 0}^{\frac{L_{\text{Trotter}}} {\texttt{NWrap}}-1}       \underbrace{\prod_{\tau=1}^{\texttt{NWrap}} \bm{B}_{i  \cdot  \texttt{NWrap}+ \tau} }_{ \equiv \mathcal{\bm{B}}_i}\right)^{-1}.
\end{equation}
Within the auxiliary field QMC implementation of the ALF project, we are by default employing
the strategy of forming a product of QR-decompositions which was proven to be weakly backwards stable in Ref.~\cite{Bai2011}.
The key idea is to efficiently separate the scales of a matrix from the orthogonal part of a matrix.
This can be achieved using a QR decomposition of a matrix $\bm{A}$ in the form $\bm{A}_i = \bm{Q}_i \bm{R}_i$. The matrix $\bm{Q}_i$ is unitary and hence in the usual $2$-norm it holds that $\kappa_2(\bm{Q}_i) = 1$.
To get a handle on the condition number of $\bm{R}_i$ we will form the
diagonal matrix
\begin{equation}
(\bm{D}_i)_{n,n} = |(\bm{R}_i)_{n,n}|
\label{eq:diagnorm}
\end{equation}
and set $\tilde{\bm{R}}_i = \bm{D}_i^{-1} \bm{R}_i$
This gives the decomposition
\begin{equation}
\bm{A}_i = \bm{Q}_i \bm{D}_i \tilde{\bm{R}}_i.
\end{equation}
$\bm{D}_i$ now contains the row norms of the original $\bm{R}_i$ matrix and hence attempts to separate off the total scales of the problem from $\bm{R}_i$.
This is similar in spirit to the so-called matrix equilibration which tries to improve the condition number of a matrix by suitably chosen column and row scalings.
Due to a theorem by van der Sluis \cite{vanderSluis1969} we know that the choice in Eq.~\eqref{eq:diagnorm} is almost optimal among all diagonal matrices $\bm{D}$ from the space of diagonal matrices 
$\mathcal{D}$ in the sense that
\begin{equation*}
\kappa_p((\bm{D}_i)^{-1} \bm{R}_i ) \leq n^{1/p} \min_{\bm{D} \in \mathcal{D}} \kappa_p(\bm{D}^{-1} \bm{R}_i).
\end{equation*}
Now, given an initial decomposition of $\bm{A}_{j-1} = \prod_i \mathcal{\bm{B}}_i = \bm{Q}_{j-1} \bm{D}_{j-1} \bm{T}_{j-1}$ an update
$\mathcal{\bm{B}}_j \bm{A}_{j-1}$ is formed in the following three steps:
\begin{enumerate}
\item Form $ \bm{M}_j = (\mathcal{\bm{B}}_j \bm{Q}_{j-1}) \bm{D}_{j-1}$. Note the parentheses.
\item Do a QR decomposition of $\bm{M}_j = \bm{Q}_j \bm{D}_j \bm{R}_j$. This gives the final $\bm{Q}_j$ and $\bm{D}_j$.
\item Form the updated $\bm{T}$ matrices $\bm{T}_j = \bm{R}_j \bm{T}_{j-1}$.
\end{enumerate}
While this might seem like quite an effort that has to be performed for every multiplication
it has to be noted that even with this stabilization scheme the algorithm preserves the time complexity class of $\mathcal{O}(\beta N^3)$ expressed in the physical parameters inverse temperature $\beta$ and lattice size $N$.
While there is no analytical expression for the dependence of the stability on the physical parameters
our experience has been that for a given number of stabilization steps along the imaginary time axis [in the notation of Eq.~(\ref{eqn:green_stable}) this number is $L_{\text{Trotter}}/\texttt{NWrap}$], the precision will be largely invariant of the system size $N$,
whereas with increasing inverse temperature $\beta$ the number of stabilization steps often has to be increased to maintain a given precision.
The effectiveness of the stabilization \emph{has} to be judged for every simulation from the output file \path{info} (Sec.~\ref{sec:output_prec}). For most simulations there are two values to look out for:
\begin{itemize}
\item \texttt{Precision Green}
\item \texttt{Precision Phase}
\end{itemize}
The Green function as well as the average phase are usually numbers with a magnitude of $\mathcal{O} (1)$. 
For that reason we recommend that \path{NWrap} is chosen such that the mean precision is of the order of $10^{-8}$ or better.  
We have included typical values of \texttt{Precision Phase} and of the mean and the maximal values of \texttt{Precision Green} in the 
discussion of example simulations, see Sec.~\ref{sec:prec_charge} and Sec.~\ref{sec:prec_spin}.
%
\subsection{Monte Carlo sampling}\label{sec:sampling}
%
Error estimates  in Monte Carlo simulations  can be  delicate and are based on the central limit theorem \cite{Negele}. This theorem requires independent 
measurements and  a finite variance.
In this subsection we will give examples of the issues that a user will have to look out for while 
using a Monte Carlo code. Those effects are part of the common lore of the field
and we can only touch on them briefly  in this text.
For a deeper understanding of the inherent issues of Markov chain Monte Carlo methods 
we refer the reader to the pedagogical introduction in chapter 1.3.5 of Krauth\cite{Krauth2006}, the overview article of Sokal~\cite{Sokal89},  the more specialized literature by Geyer~\cite{Geyer1992} and chapter 6.3 of Neal~\cite{neal1993}. 

In general, one distinguishes local from global updates. As the name suggest, the local update corresponds to a small change of the configuration, e.g. a single spin flip of one of the $L_{\mathrm{Trotter}}(M_I+M_V)$ field entries (see Sec.~\ref{sec:updating}), whereas a global update changes a significant part of the configuration. The default update scheme of the implementation at hand are local updates such that a minimum amount of moves is required to generate a independent configuration. The associated time scale is called  the autocorrelation time, $T_\mathrm{auto}$, and is generically dependent upon the choice of the observables. 

 Our unit of \textit{sweeps} is defined such that each field is visited twice in a sequential propagation from $\tau = 0$ to $\tau = L_{\text{ Trotter}}$  and back.  A single sweep will  generically not  suffice to produce an independent  configuration.
In fact, the autocorrelation time $T_\mathrm{auto}$ characterizes the required time scale to generate an independent  values of $\langle\langle\hat{O}\rangle\rangle_C$ for the observable $O$. This has several consequences for the Monte Carlo simulation:
\begin{itemize}
	\item First of all, we start from a randomly chosen field configuration such that one has to invest \textit{at least}  one, but generically much more, $T_\mathrm{auto}$ to generate relevant, equilibrated configurations before reliable measurements are possible. This phase of the simulation is known as the warm-up or burn-in phase. In order to keep the code as flexible as possible (different simulations might have different autocorrelation times), measurements are taken from the very beginning. Instead, we provide the parameter \path{n_skip} for the analysis to ignore the first \path{n_skip} bins.
	\item Secondly, our implementation averages over a given amount of measurements   set by the variable \texttt{NSWEEPS}  before storing the results, known as one bin, on the disk.  A bin corresponds to \texttt{NSWEEPS}  sweeps. The  error analysis requires statistically  independent bins to generate reliable confidence estimates. If bins are to small (averaged over a period shorter then $T_\mathrm{auto}$), the error bars are then typically underestimated. Most of the time, the autocorrelation time is unknown before the simulation is started.  Sometimes the used compute cluster does not allow single runs long enough to generate appropriately sized bins. Therefore, we provide the \path{N_rebin} parameter that specifies how many bins are combined into a new bin during the error analysis. In general, one should check that a further increase of the bin size does not change the error estimate   (For an explicit example, the reader is referred to Sec.~\ref{sec:autocorr} and the appendix of Ref.~\cite{Assaad02}).

The \path{N_rebin} variable can be used to control a second issue. The distribution of the Monte Carlo estimates $\langle\langle\hat{O}\rangle\rangle_C$ is unknown. The result in the form $(\mathrm{mean}\pm \mathrm{error})$ assumes a Gaussian distribution. Every original distribution with a finite variance turns into a Gaussian one, once it is folded often enough (central limit theorem). Due to the internal averaging (folding) within one bin, many observables are already quite Gaussian. Otherwise one can increase \path{N_rebin} further, even if the bins are already independent~\cite{Bercx17}.
	\item The third issue concerns time displaced correlation functions. Even if the configurations are independent, the fields within the configuration are still correlated. Hence, the data for $S_{\alpha,\beta}(\vec{k},\tau)$ (see Sec.~\ref{sec:obs}; Eqn.~\ref{eqn:s}) and $S_{\alpha,\beta}(\vec{k},\tau+\Delta\tau)$ are also correlated. Setting the switch \path{N_Cov = 1} triggers the calculation of the covariance matrix in addition to the usual error analysis. The covariance is defined by
	\begin{equation}
		COV_{\tau \tau'}=\frac{1}{N_{\text{Bin}}}\left\langle\left(S_{\alpha,\beta}(\vec{k},\tau)-\langle S_{\alpha,\beta}(\vec{k},\tau)\rangle\right)\left(S_{\alpha,\beta}(\vec{k},\tau')-\langle S_{\alpha,\beta}(\vec{k},\tau')\rangle\right)\right\rangle\,.
	\end{equation}
An example where this information is necessary is the  calculation of mass gaps extracted by fitting the  tail  of the time displaced correlation function.  Omitting  the covariance matrix will  underestimate the  error.
\end{itemize}

%
\subsubsection{The Jackknife resampling method}\label{sec:jack}
%
For each observable $\hat{A}, \hat{B},\hat{C} \cdots$ the Monte Carlo program computes a data set of $N_{\text{Bin}}$ (ideally) independent values where for each observable the measurements belong to the same  statistical distribution.  In the general case, we would like to evaluate a function of expectation values, $f(\langle \hat{A} \rangle, \langle \hat{B} \rangle, \langle \hat{C} \rangle  \cdots)$ --
see for example the expression (\ref{eqn:obs_rw}) for the observable including reweighting --
and are interested in the statistical estimates of its mean value  and the standard error of the mean.
A numerical method for the statistical analysis of a given function $f$ which properly handles error propagation and correlations among the observables is the Jackknife method, which is, like the related Bootstrap method, a resampling scheme \cite{efron1981}.
Here we briefly review the \textit{delete-1 Jackknife} scheme which is based on the idea to generate $N_{\text{bin}}$ new data sets of size $N_{\text{bin}}-1$ by consecutively removing one data value from the original set. By $A_{(i)}$ we denote the arithmetic mean for the observable $\hat{A}$, without the $i$-th data value $A_{i}$, namely
\begin{equation}
A_{(i)} \equiv \frac{1}{N_{\text{Bin}}-1} \sum\limits_{k=1,\,k\neq i}^{N_{\text{Bin}}} A_{k}\;.
\end{equation}
As the corresponding quantity for  the function $f(\langle \hat{A} \rangle, \langle \hat{B} \rangle, \langle \hat{C} \rangle  \cdots)$, we define 
\begin{equation}
f_{(i)}(\langle \hat{A} \rangle, \langle \hat{B} \rangle, \langle \hat{C} \rangle  \cdots) \equiv
f( A_{(i)}, B_{(i)},C_{(i)}\cdots)\;.
\end{equation}
Following the convention in the literature, we will denote the final Jackknife estimate of the mean by $f_{(\cdot)}$ and its standard error by $\Delta f$. The Jackknife mean is  given by
\begin{equation}
\label{eqn:jack_mean}
f_{(\cdot)}(\langle \hat{A} \rangle, \langle \hat{B} \rangle, \langle \hat{C} \rangle  \cdots) =
\frac{1}{N_{\text{Bin}}}\sum\limits_{i=1}^{N_{\text{Bin}}} f_{(i)}(\langle \hat{A} \rangle, \langle \hat{B} \rangle, \langle \hat{C} \rangle  \cdots)\;,
\end{equation}
and the standard error, including bias correction, is given by
\begin{equation}
\label{eqn:jack_error}
(\Delta f)^{2} = 
\frac{N_{\text{Bin}}-1}{N_{\text{Bin}}} \sum\limits_{i=1}^{N_{\text{Bin}}}
\left[f_{(i)}(\langle \hat{A} \rangle, \langle \hat{B} \rangle, \langle \hat{C} \rangle  \cdots)
- f_{(\cdot)}(\langle \hat{A} \rangle, \langle \hat{B} \rangle, \langle \hat{C} \rangle  \cdots)\right]^{2}\;.
\end{equation}
In case of $f=\langle\hat A\rangle$, the results (\ref{eqn:jack_mean}) and (\ref{eqn:jack_error}) reduce to the plain sample average and the standard, bias corrected, estimate of the error.

%
\subsubsection{An explicit example of error estimation}\label{sec:autocorr}
%
In the following we use one of our examples, the Hubbard model on a square lattice in the $M_z$ Hubbard-Stratonovich decoupling (see Sec.~\ref{sec:walk1.1}), to show explicitly how to estimate errors.  We will equally show that the  autocorrelation time is dependent upon the  choice of the observable.  In fact, different observables within the same run can have different autocorrelation times  and of course, this time scale depends on the  parameter choice.  Hence, the user has to check  autocorrelations of individual observables for each simulation!  Typical regions of the phase diagram that require special attention are critical points  where length scales diverge.  

To determine the autocorrelation time, we calculate the correlation function
\begin{equation}
\label{eqn:autocorrel}
	Auto_{\hat{O}}(t_{\textrm{QMC}})=\sum_{i=0}^{N_{\textrm{Bin}}-t_{\textrm{QMC}}}\frac{\left(O_i-\left\langle \hat{O}\right\rangle \right)\left(O_{i+t_{\textrm{QMC}}}-\left\langle \hat{O}\right\rangle \right)}{\left(O_i-\left\langle \hat{O}\right\rangle \right)\left(O_{i}-\left\langle \hat{O}\right\rangle \right)}\, ,
\end{equation}
where $O_i$ refers to the Monte Carlo estimate of the observable $\hat{O}$ in the $i^{\text{th}}$ bin. This function typically shows an exponential decay and the decay rate defines the autocorrelation time.
\begin{figure}
\includegraphics[width=1.0\textwidth]{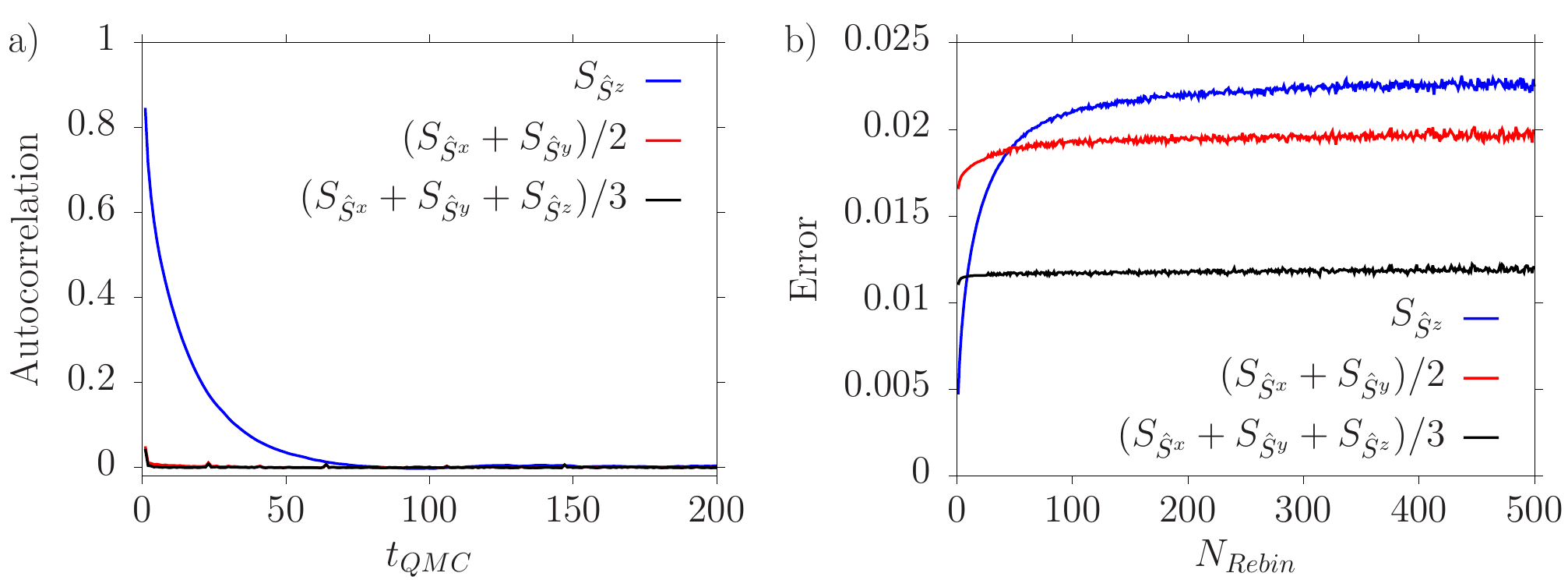}
	\caption{\label{fig_autocorr}
	        The autocorrelation function $Auto_{\hat{O}}(t_{\textrm{QMC}})$ (a) and the scaling of the error with effective bin size (b) of three equal time spin-spin correlation functions $\hat{O}$ of the Hubbard model in the $M_z$ decoupling (see Sec.~\ref{sec:walk1.1}). Simulations were done on a $ 6 \times 6$ square lattice, with  $U/t=4$ and $\beta t = 6$.  The original bin contained only one sweep and we calculated around one million bins on a single core. The different  autocorrelation times for the $xy$-plane compared to the $z$-direction can be detected from the decay rate of the autocorrelation function (a) and from the point where saturation of the error sets in (b), which defines the required effective bin size for independent measurements. Apparently and as argued in the text, the  improved estimator $(S_{\hat{S}^{x}} + S_{\hat{S}^{y}}+ S_{\hat{S}^{z}})/3$  has the smallest autocorrelation time.
 }
\end{figure}
Figure~\ref{fig_autocorr} (a) shows the autocorrelation functions $Auto_{\hat{O}}(t_{\textrm{QMC}})$ for three spin-spin-correlation functions [Eq.~(\ref{eqn:s})] at momentum $\vec{k}=(\pi,\pi)$ and at $\tau=0$: 

$\hat{O} = S_{\hat{S}^{z}}$ for the $z$ spin direction, 
$\hat{O} =(S_{\hat{S}^{x}} + S_{\hat{S}^{y}})/2$ for the $xy$ plane, and
$\hat{O} =(S_{\hat{S}^{x}} + S_{\hat{S}^{y}}+ S_{\hat{S}^{z}})/3$ for the total spin.
 The Hubbard model  has a $SU(2)$ spin symmetry. However, we chose a HS field which couples to the $z$-component of the magnetization,  $M_z$,  such that each configuration breaks this symmetry. Of course, after Monte Carlo averaging one expects restoration of the symmetry. The model, on bipartite  lattices,  shows spontaneous spin-symmetry breaking at $T=0$ and in the thermodynamic limit.  At finite temperatures, and within the so-called renormalized classical regime,  quantum antiferromagnets have a length scale  that  diverges  exponentially  with decreasing temperatures \cite{Chakravarty88}.     
The parameter set chosen for Fig.~\ref{fig_autocorr}  is non-trivial in the sense that it places the Hubbard model in this renormalized classical regime where the correlation length is substantial.  Figure~\ref{fig_autocorr}  clearly shows a very short autocorrelation time for the $xy$-plane whereas we detect a considerably longer  autocorrelation time  for the $z$-direction.  This is a direct consequence of the {\it long} magnetic length scale and the chosen decoupling.
The physical reason for the long autocorrelation time  corresponds to  the restoration of the $SU(2)$ spin symmetry.    This insight can be used to define an improved, $SU(2)$ symmetric estimator for the spin-spin correlation function, namely
$(S_{\hat{S}^{x}} + S_{\hat{S}^{y}} + S_{\hat{S}^{z}})/3$. 
 Thereby, global spin rotations are no longer an issue and this improved estimator  shows the shortest autocorrelation time as seen clearly in Fig.~\ref{fig_autocorr} (b). Other ways to tackle large autocorrelation can be global updates or parallel tempering.

Using the time series of Monte Carlo samples we would like to obtain estimates of the mean and the standard error of the mean.
A simple method which we will describe in this tutorial is the rebinning method, also known in the literature as rebatching, where a fixed number (denoted by \path{N_rebin}) of adjacent original bins are aggregated to form a new effective bin.
In addition to measuring the decay rate of the autocorrelation function (\ref{eqn:autocorrel}), a measure for the autocorrelation time  can be also obtained by the rebinning method. 
For a comparison to other methods of estimating the autocorrelation time we refer the reader to the literature \cite{Thompson2010, Geyer1992, neal1993}.
A reliable error analysis requires independent bins, otherwise the error is typically underestimated. This behavior is observed in Fig.~\ref{fig_autocorr} (b), where the effective bin size has been systematically increased by rebinning. If the effective bin size is smaller than the autocorrelation time the error will be underestimated. When the effective bin size becomes  larger than the autocorrelation time converging behavior sets in and in this region the error estimate will be correct.

For the analysis of the Monte Carlo data (see Sec.~\ref{sec:analysis}), the user can provide a finite value for \path{N_auto} to trigger the computation of  autocorrelation functions $Auto_{\hat{O}}(t_{\textrm{QMC}})$ in the range $t_{\text{QMC}}=[0,\textrm{\path{N_auto}}]$. 
Since these computations are quite time consuming and require many Monte Carlo bins the default value is  \path{N_auto=0} if unspecified. To produce Fig.~\ref{fig_autocorr}, we set $\textrm{\path{N_auto}}=500$ and used a total of approximately one million bins.

%
\subsection{Pseudo code description}\label{sec:pseudocode}
%
\begin{algorithm}[H]
\caption{Basic structure of the auxiliary field QMC implementation in \texttt{Prog/main.f90}}
\label{alg:1}
\begin{algorithmic}[1]

\State{\textbf{call} ham\_set}\Comment{\textit{Set the Hamiltonian and the lattice}}

\State{\textbf{call} confin}\Comment{\textit{Read in an auxiliary-field configuration or generate it randomly}}
\item[]

\For{$n=L_{\text{Trotter}}$ to $1$} \Comment{\textit{Fill the storage, needed for the first actual Monte Carlo sweep}}
\State{\textbf{call} wrapul}\Comment{\textit{Compute propagation matrices and store them at stabilization points}}
\EndFor
\item[]

\For{$n_{\text{bc}}=1$ to $N_{\text{bin}}$}\Comment{\textit{Loop over bins. The bin defines the unit of Monte Carlo time}}
\item[]

\For{$n_{\text{sw}}=1$ to $N_{\text{sweep}}$}\Comment{\textit{Loop over sweeps. Each sweep updates twice}}
\Statex\Comment{\textit{(upward and downward in imag. time) the space-time lattice of auxiliary fields}}
\item[]

\For{$n_{\tau}=1$ to $L_{\text{Trotter}}$}\Comment{\textit{Upward sweep}}
\State{\textbf{call} wrapgrup}\Comment{\textit{Propagate Green fct. from $n_{tau}-1$ to $n_{\tau}$, and compute new}}
\Statex\Comment{\textit{estimate of Green fct. at $n_{\tau}$, using sequential updates}}
\item[]

\ThreeIndentedLeftComment{Stabilization:}
\If{$n_{\tau}$ = stabilization point in imaginary time}

\State{\textbf{call} wrapur}\Comment{\textit{Compute propagation from previous stabilization point to $n_{\tau}$}}

\FourIndentedLeftComment{Storage management:}
\FourIndentedLeftComment{Read from storage: propagation from $L_{\text{Trotter}}$ to $n_{\tau}$}
\FourIndentedLeftComment{Write to storage : the just computed propagation}

\State{\textbf{call} cgr} \Comment{\textit{Recalculate the Green function at time $n_{\tau}$ in a stable way}}
\State{\textbf{call} control\_precisionG}\Comment{\textit{Compare propagated and recalculated Green fct.}}
\EndIf
\item[]

\If{$n_{\tau} \in [LOBS\_ST, LOBS\_EN]$}\Comment{\textit{Measure the equal time observables}}
\State{\textbf{call} obser}
\EndIf
\EndFor
\item[]

\For{$n_{\tau}=L_{\text{Trotter}}$ to $1$}\Comment{\textit{Downward sweep}}
\ThreeIndentedLeftComment{Repeat the above steps (update, propagation, stabilization, equal time}
\ThreeIndentedLeftComment{measurements) for the downward direction in imaginary time}
\EndFor
\item[]

\If{$n_{\tau} = 1$} \Comment{\textit{Measure the time displaced observables}}
\State{\textbf{call} tau\_m}
\EndIf

\EndFor

\State{\textbf{call} pr\_obs} \Comment{\textit{Calculate measurement averages for current bin and write them to disk}}
\State{\textbf{call} confout}\Comment{\textit{Write auxiliary-field configuration to disk}}

\EndFor

\end{algorithmic}
\end{algorithm}

%
\section{Data Structures and Input/Output}\label{sec:imp}
%
\subsection{Implementation of the Hamiltonian and the lattice} 
%
The module \path{Hamiltonian}, contained in the file \path{Hamiltonian.f90}, defines the model Hamiltonian, the lattice under consideration and the desired observables (Table~\ref{table:hamiltonian}). 
We have collected a number of example Hamiltonians, lattices and observables in the file  \path{Hamiltonian_Examples.f90}.  The examples are described in Sec.~\ref{sec:ex}.
To implement a user-defined model, only the module \path{Hamiltonian} has to be set up. Accordingly, this documentation focusses almost entirely  on this module and the subprograms it includes. 
The remaining parts of the code may hence be treated as a black box.  

To specify the Hamiltonian, one needs  an  \path{Operator} and a \path{Lattice} type as well as a type for the observables. These three data structures will be described in the following sections.

\begin{table}[h]
    \begin{tabular}{@{} l l l @{}}\toprule
    Subprogram & Description & Section \\\midrule
    \hlgray{\texttt{Ham\_Set}}  & Reads in model and lattice parameters from the file \texttt{parameters}\\ & and it sets the Hamiltonian by calling \texttt{Ham\_latt}, \texttt{Ham\_hop},\\ & and \texttt{Ham\_V}. & \\
    \hlgray{\texttt{Ham\_hop}}  & Sets the hopping term  $\hat{\mathcal{H}}_{T}$ by calling \texttt{Op\_make} and \texttt{Op\_set}. & \ref{sec:op}, \ref{sec:specific}\\
    \hlgray{\texttt{Ham\_V}}    & Sets the interaction terms  $\hat{\mathcal{H}}_{V}$ and $\hat{\mathcal{H}}_{I}$  by calling \texttt{Op\_make}\\ & and \texttt{Op\_set}.& \ref{sec:op}, \ref{sec:specific}\\  
    \hlgray{\texttt{Ham\_Latt}} & Sets the lattice by calling \texttt{Make\_Lattice}.& \ref{sec:latt}\\
    \hlgray{\texttt{S0}}        & A function which returns an update ratio for the Ising term \\ & $\hat{\mathcal{H}}_{I,0}$. 
    & \ref{sec:s0} \\
    \hlgray{\texttt{Alloc\_obs}} & Assigns memory storage to the observables & \\
    \hlgray{\texttt{Obser}}      & Computes the scalar observables and equal-time correlation\\ & functions. & \ref{sec:obs} \\
    \hlgray{\texttt{ObserT}}     & Computes time-displaced correlation functions. & \ref{sec:obs}\\
    \texttt{Init\_obs}  & Initializes the observables to zero. & \\    
    \texttt{Pr\_obs}    & Writes the observables to the disk by calling \texttt{Print\_bin}. \\\bottomrule    
   \end{tabular}
   \caption{Overview of the subprograms of the  module \texttt{Hamiltonian} to define the Hamiltonian, the lattice and the observables. 
   The \hlgray{highlighted} subroutines have to be modified by the user.
    \label{table:hamiltonian}}
\end{table}
%
\subsubsection{The \texttt{Operator} type}\label{sec:op}
%
The fundamental data structure in the code is the data structure \path{Operator}. It is implemented as a Fortran derived data type.
This type is used to define the Hamiltonian (\ref{eqn:general_ham}).
In general, the matrices $\textbf{T}^{(ks)}$, $\textbf{V}^{(ks)}$ and $\textbf{I}^{(ks)}$ are sparse Hermitian matrices.
Consider the  matrix   ${\bm X}$ of dimension  $N_{\mathrm{dim}} \times N_{\mathrm{dim}}$, as a representative for each of the above three matrices. 
Let us  denote  with  $ \left\{z_{1},\cdots,  z_{N}  \right\}$  a subset  of $N$ indices,
for which
\begin{equation}
X_{x,y} 
\left\{\begin{matrix}  \neq 0  &  \text{ if }   x,  y  \in \left\{ z_1, \cdots z_N \right\}\\ 
                                 = 0         &  \text{ otherwise } 
      \end{matrix}\right.
\end{equation}
Usually, we have $N\ll N_{\text{dim}}$.
 We define the $N \times N_{\mathrm{dim}}$ matrices $\mathbf{P}$  as
\begin{equation}
P_{i,x}=\delta_{z_{i},x}\;,
\end{equation}
where $i \in [1,\cdots, N ]$ and $ x  \in [1,\cdots, N_{\mathrm{dim}}]$. The matrix  $\bm{P}$ selects the non-vanishing entries of $\bm{X}$, 
which are contained in the rank-$N$  matrix $\bm{O}$:
\begin{equation}\label{eqn:xeqpdop}
\bm{X} =\bm{P}^{T} \bm{O} \bm{P}\;,
\end{equation}
and 
\begin{equation}
X_{x,y} = \sum\limits_{i,j}^{N}  P_{i,x}  O_{i,j} P_{j,y}=\sum\limits_{i,j}^{N} \delta_{z_{i},x}  O_{ij} \delta_{z_{j},y} \;.
\end{equation}
Since  the  $\bm{P}$ matrices have only one non-vanishing entry per column,  they can conveniently be stored as a vector $\vec{P}$, with entries
\begin{equation}
     P_i = z_i.
\end{equation}  
There are  many useful  identities which emerge from this  structure. For example: 
\begin{equation}
	e^{\bm{X}} =  e^{\bm{P}^{T} \bm{O} \bm{P}}   = \sum_{n=0}^{\infty}  \frac{\left( \bm{P}^{T} \bm{O} \bm{P} \right)^n}{n!} = \mathds{1}+ \bm{P}^{T} \left(e^{ \bm{O} }-\mathds{1} \right) \bm{P}\;,
\end{equation}
since 
\begin{equation} 
	 \bm{P} \bm{P}^{T}= \mathds{1}_{N\times N}.
\end{equation}

In the code, we define a structure called \path{Operator} to capture the above. 
This type \path{Operator} bundles several components that are needed to define and use an operator matrix in the program.  
%
\subsubsection{Specification of the model}\label{sec:specific}
%
\begin{table}[h]
    \begin{tabular}{@{} l l l @{}}\toprule
    Variable & Type & Description \\\midrule
    \hlgray{\texttt{Op\_X\%N}}       & Integer     &  Effective dimension $N$ \\
    \hlgray{\texttt{Op\_X\%O}}       & Complex    &  Matrix  $\mathbf{O}$  of dimension $N \times N$\\
    \hlgray{\texttt{Op\_X\%P}}       & Integer   &  Matrix $\mathbf{P}$  encoded as a vector of dimension $N$\\
    \hlgray{\texttt{Op\_X\%g}}       & Complex    &  Coupling strength $g$ \\  
    \hlgray{\texttt{Op\_X\%alpha}}   & Complex  &  Constant $\alpha$ \\
    \hlgray{\texttt{Op\_X\%type}}    & Integer   &  Parameter to set the type of 
                                             HS transformation\\
                             &   &  (1 = Ising, 2 = discrete HS for perfect-square term)  \\ 
    \texttt{Op\_X\%U}            & Complex &  Matrix containing the eigenvectors of $\mathbf{O}$  \\
    \texttt{Op\_X\%E}            & Real &  Eigenvalues of $\mathbf{O}$ \\
    \texttt{Op\_X\%N\_non\_zero} & Integer &  Number of non-vanishing eigenvalues of $\mathbf{O}$ \\\bottomrule
   \end{tabular}
   \caption{Member variables of the \texttt{Operator}  type. 
   In the left column, the letter \texttt{X} is a placeholder for the letters \texttt{T} and \texttt{V}, 
   indicating hopping and interaction operators, respectively.
   The \hlgray{highlighted} variables have to be specified by the user.
    \label{table:operator}}
\end{table}
In this section we show how to specify the  Hamiltonian (\ref{eqn:general_ham}) in the code. 
More precisely, we have to set the matrix representation of the imaginary-time propagators --
$ e^{-\Delta \tau {\bm T}^{(ks)}}$, $e^{  \sqrt{ -\Delta \tau  U_k} \eta_{k\tau} {\bm V}^{(ks)} }$, and $e^{  -\Delta \tau s_{k\tau}  {\bm I}^{(ks)}}$ -- that appear in the 
partition function (\ref{eqn:partition_2}).  For each pair of indices $(k,s)$, these terms have the general form
\begin{equation}\label{eqn:exponent_mat}
\text{Matrix Exponential}=
e^{g \,\phi(\texttt{type})\,\mathbf{X} }\;.
\end{equation}
In case of the  perfect-square term,  we additionally have to set the constant $\alpha$, see the definition of the operators $\hat{V}^{(k)}$ in Eq.~(\ref{eqn:general_ham_v}).
The data structures which hold all the above information are variables of the type \path{Operator} (see Table \ref{table:operator}). 
For each pair of indices $(k,s)$, we store the following parameters in an \path{Operator} variable:
\begin{itemize}
\item $\vec{P}$ and   $ \mathbf{O}$   defining the matrix $\mathbf{X}$ [see Eq.~(\ref{eqn:xeqpdop})]
\item the constants $g$, $\alpha$
\item optionally: the type \texttt{type} of the discrete fields $\phi$
\end{itemize}
In case of the Ising term,  we store \path{type=1} which sets $\phi_{k\tau}=s_{k\tau}$. 
In case of the perfect-square term, the field results from the discrete HS transformation (\ref{HS_squares}) and we store \texttt{type=2} which sets $\phi_{k\tau}=\eta_{k\tau}$. 
Note that we have dropped the color index $\sigma$, since the implementation uses the $SU(N_{\mathrm{col}})$ invariance of the Hamiltonian. 

Accordingly, the following data structures fully describe the  Hamiltonian (\ref{eqn:general_ham}):
\begin{itemize}
\item For the hopping Hamiltonian (\ref{eqn:general_ham_t}), we have to set the exponentiated hopping matrices $ e^{-\Delta \tau {\bm T}^{(ks)}}$: 

In this case $\mathbf{X}^{(ks)}=\mathbf{T}^{(ks)}$. Precisely, a single variable  \texttt{Op\_T}  describes the operator matrix
\begin{equation}
            \left( \sum_{x,y}^{N_{\mathrm{dim}}} \hat{c}^{\dagger}_x T_{xy}^{(ks)} \hat{c}^{\phantom{\dagger}}_{y}  \right)  \;,
\end{equation} 
where $k=[1, M_{T}]$ and $s=[1, N_{\mathrm{fl}}]$.  
To make contact with the general expression (\ref{eqn:exponent_mat}) we set $g=-\Delta \tau$ (and $\alpha = 0$).
In case of the hopping matrix, the type variable $\texttt{Op\_T\%type}$  is neglected by the code. 
All in all, the corresponding array of structure variables is  \texttt{Op\_T(M$_T$,N$_{fl}$)}.

\item For the interaction Hamiltonian (\ref{eqn:general_ham_v}), which is of perfect-square type, we have to set the exponentiated matrices $e^{  \sqrt{ -  \Delta \tau  U_k} \eta_{k\tau} {\bm V}^{(ks)} }$:

In this case, ${\mathbf X}  = \mathbf{V}^{(ks)}$. A single variable  \texttt{Op\_V}  describes the operator matrix:
\begin{equation}
             \left[ \left( \sum_{x,y}^{N_{\mathrm{dim}}} \hat{c}^{\dagger}_x V_{x,y}^{(ks)} \hat{c}^{\phantom{\dagger}}_{y}  \right)  + \alpha_{ks} \right]  \;,
\end{equation} 
where $k=[1, M_{V}]$ and $s=[1, N_{\mathrm{fl}}]$. 
To make contact with the general expression (\ref{eqn:exponent_mat}) and to set the constant $\alpha$, we choose $g = \sqrt{-\Delta \tau  U_k}$ and  $\alpha = \alpha_{ks}$. 
The discrete Hubbard-Stratonovich decomposition which is used for the perfect-square interaction, is selected by setting the type variable to $\texttt{Op\_V\%type}=2$.
All in all, the required structure variables \texttt{Op\_V} are defined  using the array \texttt{Op\_V(M$_V$,N$_{fl}$)}.

\item For the Ising interaction Hamiltonian (\ref{eqn:general_ham_i}), we have to set the exponentiated matrices $e^{  -\Delta \tau s_{k\tau}  {\bm I}^{(ks)}}$:

In this case, $\bm{X}  = \bm{I}^{(k,s)} $.  
A single variable  \texttt{Op\_V} then  describes the operator matrix:
\begin{equation}
            \left( \sum_{x,y}^{N_{\mathrm{dim}}} \hat{c}^{\dagger}_x I_{xy}^{(ks)} \hat{c}^{\phantom{\dagger}}_{y}  \right)  \;,
\end{equation} 
where $k=[1, M_{I}]$ and $s=[1, N_{\mathrm{fl}}]$. 
To make contact with the general expression (\ref{eqn:exponent_mat}), we set  $g = -\Delta \tau$ (and $\alpha = 0$).
The Ising interaction is specified by setting the type variable  $\texttt{Op\_V\%type=1}$. 
All in all, the required structure variables are contained in the array \texttt{Op\_V(M$_{I}$,N$_{fl}$)}.

\item In case of a full interaction [perfect-square term (\ref{eqn:general_ham_v}) and Ising term (\ref{eqn:general_ham_i})],
we  define  the corresponding doubled array \texttt{Op\_V(M$_V$+M$_I$,N$_{fl}$) } and set the variables separately for both ranges of the array according to the above.  

\end{itemize}
%
\subsubsection{The \texttt{Lattice} type}\label{sec:latt}
%
We have a lattice module  which can generate one- and two-dimensional Bravais lattices.
Note that the orbital structure of each unit cell has to be specified by the user in the Hamiltonian module. 
 The user has to specify unit vectors $\vec{a}_1$ and $\vec{a}_2$ as well as the size of the  lattice. The size is  characterized by  two vectors $\vec{L}_1$ and $\vec{L}_2$   and  the lattice is placed on a torus 
 (periodic boundary conditions): 
\begin{equation}
	\hat{c}_{\vec{i} + \vec{L}_1 }  = \hat{c}_{\vec{i} + \vec{L}_2 }  = \hat{c}_{\vec{i}}.
\end{equation}
The function call 
\lstset{style=fortran}
\begin{lstlisting} 
Call Make_Lattice( L1, L2, a1,  a2, Latt )
\end{lstlisting}
will generate the lattice   \texttt{Latt} of type \texttt{Lattice}.   Note again that  the orbital structure of the unit cell has to be provided by the user.    The reciprocal lattice vectors are defined by: 
\begin{equation}
\label{Latt.G.eq}
	\vec{a}_i  \cdot \vec{g}_i = 2 \pi \delta_{i,j}, 
\end{equation}
and the Brillouin zone corresponds to the Wigner-Seitz cell of the lattice. 
With $\vec{k} = \sum_{i} \alpha_i  \vec{g}_i $, the  k-space quantization follows from: 
\begin{equation}
\begin{bmatrix}
	\vec{L}_1 \cdot \vec{g}_1  &  \vec{L}_1 \cdot \vec{g}_2  \\
	\vec{L}_2  \cdot \vec{g_1} & \vec{L}_2 \cdot  \vec{g}_2  
\end{bmatrix}
\begin{bmatrix}
   \alpha_1 \\
   \alpha_2
\end{bmatrix}
=
2 \pi 
\begin{bmatrix}
   n \\
   m
\end{bmatrix}
\end{equation}
such that 
\begin{eqnarray}
\label{k.quant.eq}
     \vec{k} &=&  n \vec{b}_1  + m \vec{b}_2\;, \text{ with}\nonumber\\
     \vec{b}_1 &=& \frac{2 \pi}{ (\vec{L}_1 \cdot \vec{g}_1)  (\vec{L}_2 \cdot  \vec{g}_2 )  - (\vec{L}_1 \cdot \vec{g}_2) (\vec{L}_2  \cdot \vec{g_1} ) }   \left[  (\vec{L}_2 \cdot  \vec{g}_2) \vec{g}_1 -   (\vec{L}_2  \cdot \vec{g_1} ) \vec{g}_2 \right]\;, \text{   and  } \nonumber \\ 
      \vec{b}_2 &=& \frac{2 \pi}{ (\vec{L}_1 \cdot \vec{g}_1)  (\vec{L}_2 \cdot  \vec{g}_2 )  - (\vec{L}_1 \cdot \vec{g}_2) (\vec{L}_2  \cdot \vec{g_1} ) }   
           \left[  (\vec{L}_1 \cdot  \vec{g}_1) \vec{g}_2 -   (\vec{L}_1  \cdot \vec{g_2} ) \vec{g}_1 \right] \;.
\end{eqnarray}

\begin{table}[h]
   \begin{tabular}{@{} l l l @{}}\toprule
    Variable  & Type & Description \\\midrule
     \hlgray{\texttt{Latt\%a1\_p}, \texttt{Latt\%a2\_p}}   & Real     & Unit vectors $\vec{a}_1$,  $\vec{a}_2$ \\ 
     \hlgray{\texttt{Latt\%L1\_p}, \texttt{Latt\%L2\_p}}   & Real     & Vectors $\vec{L}_1$, $\vec{L}_2$ that define the topology of the  lattice. \\
     									  &              &  Tilted lattices are  thereby possible to implement.  \\
    \texttt{Latt\%N}                                                 &   Integer &  Number of lattice points, $N_{\text{unit cell}}$   \\
    \texttt{Latt\%list}                                               & Integer &  Maps each lattice point $i=1,\cdots, N_{\text{unit cell}}$ to a real\\ 
                                                                             &   & space vector denoting the position of the unit cell: \\
                                                                             &   & $\vec{R}_i$ = \texttt{list(i,1)} $\vec{a}_1$ +  \texttt{list(i,2)} $\vec{a}_2$  $  \equiv i_1  \vec{a}_1 + i_2  \vec{a}_2 $ \\
    \texttt{Latt\%invlist}                                        &  Integer &   \texttt{Invlist}$(i_1,i_2) = i $ \\
    \texttt{Latt\%nnlist}                                         &  Integer &   $j = \texttt{nnlist} (i, n_1, n_2) $,  $n_1, n_2 \in [-1,1] $ \\
                                                                           &              &    $\vec{R}_j = \vec{R}_i + n_1 \vec{a}_1  + n_2 \vec{a}_2 $ \\
   \texttt{Latt\%imj}                                             &   Integer  &  $ \vec{R}_{imj(i,j)}  =  \vec{R}_i -  \vec{R}_j$.        $imj, i, j \in  1,\cdots, N_{\text{unit cell}}$\\
    \texttt{Latt\%BZ1\_p}, \texttt{Latt\%BZ2\_p}  &   Real     & Reciprocal space vectors $\vec{g}_i$   (See Eq.~\ref{Latt.G.eq})\\
    \texttt{Latt\%b1\_p}, \texttt{Latt\%b1\_p}       &   Real     &  $k$-quantization (See Eq.~\ref{k.quant.eq}) \\
    \texttt{Latt\%listk}                                           &  Integer &  Maps each reciprocal lattice point $k=1,\cdots, N_{\text{unit cell}}$\\
                                                                          &    & to a reciprocal space vector\\
                                                                          &     & $\vec{k}_k= \texttt{listk(k,1)} \vec{b}_1 +  \texttt{listk(k,2)} \vec{b}_2  \equiv k_1  \vec{b}_1 +   k_2  \vec{b}_2 $\\
    \texttt{Latt\%invlistk}                                     &    Integer    &   \texttt{Invlistk}$(k_1,k_2) = k $ \\
   \texttt{Latt\%b1\_perp\_p},  \\ 
   \texttt{Latt\%b2\_perp\_p}                             &    Real         &  Orthonormal vectors to $\vec{b}_i$.  For internal use. \\\bottomrule
   \end{tabular}
   \caption{Components of the \texttt{Lattice} type for two-dimensional lattices using as example the default lattice name \texttt{Latt}.
   The \hlgray{highlighted} variables have to be specified by the user.  Other components of the \texttt{Lattice} are generated upon calling: \texttt{Call Make\_Lattice( L1, L2, a1,  a2, Latt )}. 
    \label{table:lattice}}
\end{table}

The \path{Lattice}  module equally handles  the Fourier transformation.  For example  the  subroutine  \path{Fourier_R_to_K}   carries out the  transformation: 
\begin{equation}
	S(\vec{k}, :,:,:) =  \frac{1}{N_{unit \,cell}}  \sum_{\vec{i},\vec{j}}   e^{-i \vec{k} \cdot \left( \vec{i}-\vec{j} \right)} S(\vec{i}  - \vec{j}, :,:,:)
\end{equation}
and  \path{Fourier_K_to_R}  the  inverse Fourier transform 
 \begin{equation}
	S(\vec{r}, :,:,:) =  \frac{1}{N_{unit \,cell}}  \sum_{\vec{k} \in BZ }   e^{ i \vec{k} \cdot \vec{r}} S(\vec{k}, :,:,:).
\end{equation}
In the above,   the unspecified dimensions of  the structure factor can refer  to imaginary-time  and orbital indices. 
%
\subsection{The observable types \texttt{Obser\_Vec} and \texttt{Obser\_Latt}}\label{sec:obs}
%
Our definition  of the model includes observables [Eq.~(\ref{eqn:obs_rw})]. We have defined two observable types: \texttt{Obser\_vec}  for an array of scalar observables
such as the energy, and  \texttt{Obser\_Latt}   for correlation functions that have the lattice symmetry. In the latter case, translation symmetry can be used to provide improved estimators and to reduce the size of the output.   
We also obtain improved estimators by taking measurements in the imaginary-time interval \texttt{[LOBS\_ST,LOBS\_EN]}  (see the parameter file in Sec.~\ref{sec:input}) thereby exploiting the invariance under translation in imaginary-time.
Note that the translation symmetries  in space and in time are \textit{broken} for a given  configuration $C$ but restored by the Monte Carlo sampling. 
In general, the user defines the size and  the number of bins in the parameter file, each bin having a given amount of sweeps. Within a sweep we run sequentially through the HS and Ising fields, from time slice $1$ to time slice $L_{\text{Trotter}}$ and back.  The results of each bin are written to a file  and analyzed at the end of the run.     

To accomplish the reweighting of observables (see Sec.~\ref{sec:reweight}), for each configuration the measured value of an observable is multiplied by the factors \texttt{ZS} and \texttt{ZP}:
\begin{eqnarray}
\texttt{ZS} &=& \text{sign}(C)\;,\\
\texttt{ZP} &=& \frac{e^{-S(C)}} {\Re \left[e^{-S(C)} \right]}\;.
\end{eqnarray}
They are computed from the Monte Carlo phase of a configuration,
\begin{equation}\label{eqn:phase}
	\texttt{phase}   =   \frac{e^{-S(C)}}{ \left| e^{-S(C) }\right| }\;,
\end{equation}
which is provided by the main program.
Note that each observable structure also includes the average sign [Eq.~(\ref{eqn:sign_rw})].
%
\subsubsection{Scalar observables}
%
This data type  is described in Table  \ref{table:Obser_vec} and  is useful to compute an array of  scalar observables.   Consider  a variable \texttt{Obs} of type  \texttt{Obser\_vec}.  At the beginning of each bin,  a call to  \texttt{Obser\_Vec\_Init} in the module \texttt{observables\_mod.f90}  will  set   \texttt{Obs\%N=0},   \texttt{Obs\%Ave\_sign =0}  and  \texttt{Obs\%Obs\_vec(:)=0}.  Each time the main  program calls the routine \texttt{Obser}  in the  \texttt{Hamiltonian} module,  the counter \texttt{Obs\%N}   is incremented by one, the sign  (see Eq.~\ref{Sign.eq}) is accumulated in the  variable \texttt{Obs\%Ave\_sign},  and the desired observables (multiplied by the sign and   $\frac{e^{-S(C)}} {\Re \left[e^{-S(C)} \right]}$, see Sec.~\ref{Observables.General})  are accumulated in the vector \texttt{Obs\%Obs\_vec}.  
\begin{table}[h]
   \begin{tabular}{@{} l l l l @{}}\toprule
    Variable  &  Type      &  Description &  Contribution of  \\
        &  & & configuration $C$ \\\midrule
    \texttt{Obs\%N}                       &  Int.        &   Number of measurements &\\
    \texttt{Obs\%Ave\_sign}               &  Real     &    Cumulated sign [Eq.~(\ref{eqn:sign_rw})] & $\text{sign}(C)$  \\
    \texttt{Obs\%Obs\_vec(:)}        & Compl.      &    Cumulated vector of & \\
         &       &    observables [Eq.~(\ref{eqn:obs_rw})] &
           $ \langle \langle \hat{O}(:) \rangle \rangle_{C}\frac{e^{-S(C)}} {\Re \left[e^{-S(C)} \right]} \text{ sign }(C) $ \\
     \texttt{Obs\%File\_Vec}           &  Char.    &    Name of output file  &\\\bottomrule
   \end{tabular}
   \caption{Components of the \texttt{Obser\_vec}  type.  The table lists the data included in a variable  \texttt{Obs}  of type \texttt{Obser\_vec}.  
         \label{table:Obser_vec}}
\end{table}
At the end of the bin, a call to  \texttt{Print\_bin\_Vec}   in  module \texttt{observables\_mod.f90}  will  append the result of the bin in the file  \texttt{File\_Vec}\emph{\_scal}.  Note that this subroutine will automatically append the suffix \emph{\_scal}
to the the filename \texttt{File\_Vec}.
This suffix  is important to allow automatic analysis of the data at the end of the run. 
%
\subsubsection{ Equal time and time displaced correlation functions}
%
\begin{table}[h]
   \begin{tabular}{@{} l l l l @{}}\toprule
        Variable  &  Type      &  Description &  Contribution of  \\
        &  & & configuration $C$ \\\midrule
    \texttt{Obs\%N}                       &  Int.        &   Number of measurements &  \\
    \texttt{Obs\%Ave\_sign}  
    &  Real  &    Cumulated sign [Eq.~(\ref{eqn:sign_rw})] & $\text{sign}(C)$  \\
    \texttt{Obs\%Obs\_latt}        & Compl.      &    Cumululated  correlation & \\
     $(\vec{i}-\vec{j},\tau,\alpha,\beta)$  &     &  function [Eq.~(\ref{eqn:obs_rw})] &  $ \langle \langle \hat{O}_{\vec{i},\alpha} (\tau) \hat{O}_{\vec{j},\beta} \rangle \rangle_{C} \; \frac{e^{-S(C)}} {\Re \left[e^{-S(C)} \right]}  \text{sign}(C) $ \\
     \texttt{Obs\%Obs\_latt0($\alpha$)}        & Compl.      &    Cumulated expectation &   \\
             &     &    value [Eq.~(\ref{eqn:obs_rw})] &   $ \langle \langle \hat{O}_{\vec{i},\alpha} \rangle \rangle_{C}\frac{e^{-S(C)}} {\Re \left[e^{-S(C)} \right]}  \text{ sign }(C) $ \\
     \texttt{Obs\%File\_Latt}           &  Char.    &    Name of output file  &\\\bottomrule
   \end{tabular}
   \caption{Components of the \texttt{Obser\_latt}  type.  The table lists the data included in a variable  \texttt{Obs}  of type \texttt{Obser\_latt}.
      \label{table:Obser_latt}}
\end{table}
This data type (see Table~\ref{table:Obser_latt}) is useful so as to deal with  equal time as well as imaginary-time displaced correlation functions of the form: 
\begin{equation}\label{eqn:s}
	S_{\hat{O},\alpha,\beta}(\vec{k},\tau) =   \frac{1}{N_{\text{unit cell}}} \sum_{\vec{i},\vec{j}}  e^{- \vec{k} \cdot \left( \vec{i}-\vec{j}\right) } \left( \langle \hat{O}_{\vec{i},\alpha} (\tau) \hat{O}_{\vec{j},\beta} \rangle  - 
	  \langle \hat{O}_{\vec{i},\alpha} \rangle \langle   \hat{O}_{\vec{j},\beta}  \rangle \right).
\end{equation}
Here,  translation symmetry of the Bravais lattice is explicitly taken into account. 
The correlation function splits in a correlated part $S_{\hat{O},\alpha,\beta}^{\mathrm{(corr)}}(\vec{k},\tau)$ and a background part $S_{\hat{O},\alpha,\beta}^{\mathrm{(back)}}(\vec{k})$:
\begin{eqnarray}
  S_{\hat{O},\alpha,\beta}^{\mathrm{(corr)}}(\vec{k},\tau)
  &=&
   \frac{1}{N_{\text{unit cell}}} \sum_{\vec{i},\vec{j}}  e^{- i\vec{k} \cdot \left( \vec{i}-\vec{j}\right) }  \langle \hat{O}_{\vec{i},\alpha} (\tau) \hat{O}_{\vec{j},\beta} \rangle\label{eqn:s_corr}\;,\\
         S_{\hat{O},\alpha,\beta}^{\mathrm{(back)}}(\vec{k})
  &=&
   \frac{1}{N_{\text{unit cell}}} \sum_{\vec{i},\vec{j}}  e^{- i\vec{k} \cdot \left( \vec{i}-\vec{j}\right) }  \langle \hat{O}_{\vec{i},\alpha} (\tau)\rangle \langle \hat{O}_{\vec{j},\beta} \rangle\nonumber\\
  &=& 
  N_{\text{unit cell}}\, \langle \hat{O}_{\alpha} \rangle \langle \hat{O}_{\beta} \rangle \, \delta(\vec{k})\label{eqn:s_back}\;,
\end{eqnarray}
where translation invariance in space and time has been exploited to obtain the last line. 
The background part depends only on the expectation value $\langle \hat{O}_{\alpha} \rangle$, for which we use the following estimator 
\begin{equation}\label{eqn:o}
\langle \hat{O}_{\alpha} \rangle \equiv \frac{1}{N_{\text{unit\,cell}}} \sum\limits_{\vec{i}} \langle \hat{O}_{\vec{i},\alpha} \rangle\;.
\end{equation}
Consider a variable \texttt{Obs} of type  \texttt{Obser\_latt}. At the beginning of each bin a call to the subroutine \texttt{Obser\_Latt\_Init} in the module \texttt{observables\_mod.f90}  will  initialize  the elements of \texttt{Obs} to zero.    Each time the main program calls the   \texttt{Obser} or  \texttt{ObserT} routines one accumulates the quantity $\langle \langle \hat{O}_{\vec{i},\alpha} (\tau) \hat{O}_{\vec{j},\beta} \rangle \rangle_{C} \; \frac{e^{-S(C)}} {\Re \left[e^{-S(C)} \right]}  \text{sign}(C) $    in  \texttt{Obs\%Obs\_latt($\vec{i}-\vec{j},\tau,\alpha,\beta$)} and $ \langle \langle \hat{O}_{\vec{i},\alpha} \rangle \rangle_{C}\frac{e^{-S(C)}} {\Re \left[e^{-S(C)} \right]}  \text{ sign }(C) $  in \texttt{Obs\%Obs\_latt0($\alpha$)}.   At the end of each bin, a call to \path{Print_bin_Latt} in the module  \texttt{observables\_mod.f90}   will append the result of the bin in the specified  file \texttt{Obs\%File\_Latt}.   Note that the routine  \texttt{Print\_bin\_Latt}  carries out the Fourier transformation and prints the results in $k$-space. 
We have adopted the following naming conventions.
For equal time observables,
defined by having the second dimension  of the array  \texttt{Obs\%Obs\_latt($\vec{i}-\vec{j},\tau,\alpha,\beta$)} set to unity, 
the routine \texttt{Print\_bin\_Latt}  attaches the suffix \emph{\_eq} to \texttt{Obs\%File\_Latt}.  For  time displaced correlation functions we use the suffix \emph{\_tau}.

%
\subsection{File structure}\label{sec:files}
%
\begin{table}[h]
   \begin{tabular}{@{} l l @{}}\toprule
   Directory & Description \\\midrule
   \path{Prog/} & Main program and subroutines  \\
  \path{Libraries/} & Collection of mathematical routines \\  
  \path{Analysis/} & Routines for error analysis \\
  \path{Examples/} & Example simulations for Hubbard-type models\\
  \path{Start/}   & Parameter files and scripts  \\
  \path{Documentation/} & Documentation of the QMC code.\\\bottomrule
   \end{tabular}
   \caption{Overview of the directories.\label{table:files}}
\end{table}
The code package consists of the program directories \path{Prog/}, \path{Libraries/} and \path{Analysis/}. 
The example simulations corresponding to the walkthroughs of Sec.~\ref{sec:walk1} - \ref{sec:walk2} are included in \path{Examples/}. 
The package content is summarized in Table~\ref{table:files}.
%
\subsubsection{Input files}\label{sec:input}
%
\begin{table}[h]
   \begin{tabular}{@{} l l @{}}\toprule
   File & Description \\\midrule
  \path{parameters} &  This collects input data. We can set here the parameters for\\ & the lattice,  which model, variables of the QMC process, and\\ & the error analysis.\\
  \path{seeds} & List of integer numbers to initialize the random number\\
  &  generator and to start a simulation from scratch.\\
   (\path{confin_<threadnumber>}) & (Optionally, a HS and Ising field configuration can be\\ & provided as input.)
  \\\bottomrule
   \end{tabular}
   \caption{Overview of the input files in \texttt{Start/} required for a simulation. \label{table:input}}
\end{table}
The input files are listed in Table~\ref{table:input}. To enable restarting a previous simulation (see Table~\ref{table:scripts}) or to use a given HS and Ising field configuration as input for a new simulation, the program reads in the files \path{confin_<threadnumber>} in case they are present. It goes without saying that the dimensions of the thereby defined field configuration (number of threads, lattices size, and number of time slices) have to match the corresponding values of the parameter file.
The parameter file \path{Start/parameters} has the following form --
using as an example  the $SU(2)$-symmetric Hubbard model on a square lattice (see Sec.~\ref{sec:walk1} for a detailed walkthrough):
\lstset{style=fortran}
\begin{lstlisting} 

!============================================================================
!  Variables for the Hubb program
!----------------------------------------------------------------------------
&VAR_lattice
L1 = 4                    ! Length in direction a_1
L2 = 4                    ! Length in direction a_2
Lattice_type = "Square"	  ! a_1 = (1,0),a_2=(0,1),  Norb=1, N_coord=2
!Lattice_type ="Honeycomb"! a_1 = (1,0),a_2 =(1/2,sqrt(3)/2),Norb=2,N_coord=3
Model = "Hubbard_SU2"     ! Sets Nf=1, N_sun=2. HS field couples to the
                          ! density
!Model = "Hubbard_Mz"     ! Sets Nf=2, N_sun=1. HS field couples to the 
                          ! z-component of magnetization.  
!Model="Hubbard_SU2_Ising"! Sets Nf_1, N_sun=2 and runs only for the square
                          ! lattice
                          ! Hubbard model  coupled to transverse Ising field
/
&VAR_Hubbard              ! Variables for the Hubbard model
ham_T   = 1.D0            ! Hopping parameter
ham_chem= 0.D0            ! chemical potential
ham_U   = 4.D0            ! Hubbard interaction
Beta    = 5.D0            ! inverse temperature
dtau    = 0.1D0           ! Thereby Ltrot=Beta/dtau
/

&VAR_Ising                ! Model parameters for the Ising code
Ham_xi = 1.d0             ! Only needed if Model="Hubbard_SU2_Ising"
Ham_J  = 0.2d0
Ham_h  = 2.d0
/

&VAR_QMC                  ! Variables for the QMC run
Nwrap   = 10              ! Stabilization. Green functions is computed from
                          ! scratch after each time interval  Nwrap*Dtau
NSweep  = 500             ! Number of sweeps
NBin    = 2               ! Number of bins
Ltau    = 1               ! 1 for calculation of time displ. Green functions;
                          ! 0 otherwise
LOBS_ST = 1               ! Start measurements at time slice LOBS_ST
LOBS_EN = 50              ! End   measurements at time slice LOBS_EN
CPU_MAX = 0.1             ! Code will stop after CPU_MAX hours. 
                          ! If not specified, code will stop after Nbin bins.
/                          
&VAR_errors               ! Variables for analysis programs
n_skip  = 1               ! Number of bins that will be skipped. 
N_rebin = 1               ! Rebinning  
N_Cov   = 0               ! If set to 1 covariance will be computed
                          ! for unequal time correlation functions.   
N_auto  = 100             ! If set to >0 autocorrelation function will be
                          ! computed for scalar and equal time observables.                  
/            
\end{lstlisting}
%
\subsubsection{Output: Observables} \label{sec:output_obs}
%
\begin{table}[h]
   \begin{tabular}{@{} l l @{}}\toprule
   File & Description \\\midrule
   \path{info} & After completion of the simulation, this file documents para-\\
   & meters of the model, the QMC run and simulation metrics\\
   & (precision, acceptance rate, wallclock time).\\
   \path{X_scal} & Results of equal time measurements of scalar observables. \\
   & The placeholder \path{X} stands for the observables \path{Kin, Pot, Part},\\ & and \path{Ener}. \\
   \path{Y_eq, Y_tau} & Results of equal time and time displaced measurements of cor-\\
   & relation functions. The placeholder \path{Y} stands for \path{Green, SpinZ,}\\ &  \path{SpinXY}, and \path{Den}. \\   
   \path{confout_<thread number>} & Output files for the HS and Ising field configuration. \\\bottomrule
   \end{tabular}
   \caption{Overview of the standard output files. 
  See Sec.~\ref{sec:obs} for the definitions of observables and correlation functions. \label{table:output}}
\end{table}
The standard output files are listed in Table~\ref{table:output}. 
The output of the measured data is organized in bins. One bin corresponds to the arithmetic average 
over a fixed number of individual measurements which depends 
on the chosen measurement interval \path{[LOBS_ST,LOBS_EN]} on the imaginary-time axis and on the number \path{NSweep} of Monte Carlo sweeps. If the user runs an MPI parallelized version of the code, the average also extends 
over the number of MPI threads. The formatting of the output for a single bin depends on the observable type,  \path{Obs_vec} or \path{Obs_Latt}:
\begin{itemize}
\item Observables of type \path{Obs_vec}:
For each additional bin, a single new line is added to the output file.
In case of an observable with \path{N_size} components, the formatting is 
\begin{verbatim}
N_size+1  <measured value,1> ... <measured value,N_size>  <measured sign>
\end{verbatim}
The counter variable \path{N_size+1} refers to the number of measurements per line, including the phase measurement. 
This format is required by the error analysis routine (see Sec.~\ref{sec:analysis}). 
Scalar observables like kinetic energy, potential energy, total energy and particle number are treated as a vector 
of size \path{N_size=1}.

\item Observables of type \path{Obs_Latt}:
For each additional bin, a new data block is added to the output file. 
The block consists of the expectation values [Eq.~(\ref{eqn:o})] contributing to the background part [Eq.~(\ref{eqn:s_back})] of the correlation function,
and the correlated part [Eq.~(\ref{eqn:s_corr})] of the correlation function.
For imaginary-time displaced correlation functions, the formatting of the block follows this scheme:
\begin{alltt}
<measured sign>  <N_orbital>  <N_unit_cell> <N_time_slices> <dtau>
do alpha = 1, N_orbital
    \(\langle\hat{O}\sb{\alpha}\rangle \)
enddo
do i = 1, N_unit_cell
   <reciprocal lattice vector k(i)>
   do tau = 1, N_time_slices
      do alpha = 1, N_orbital
         do beta = 1, N_orbital
            \(\langle\,S\sb{\hat{O},\alpha,\beta}\sp{(corr)}(k(i),\tau)\rangle\)
         enddo
      enddo
   enddo
enddo
\end{alltt}
The same block structure is used for equal time correlation functions, except for the entries  \path{<N_time_slices>} and \path{<dtau>} 
which are not present in the latter.
Using this structure for the bins as input,
the full correlation function $S_{\hat{O},\alpha,\beta}(\vec{k},\tau)$ [Eq.~(\ref{eqn:s})] is then calculated by calling the error analysis routine (see Sec.~\ref{sec:analysis}).
\end{itemize}

%
\subsubsection{Output: Precision} \label{sec:output_prec}
%
The finite temperature  auxiliary field QMC algorithm is known to be numerically  unstable, as discussed in Sec.~\ref{sec:stable}.
The origin the numerical instabilities arises  from the imaginary-time propagation which invariably leads to exponentially small and exponentially large scales.
Numerical stabilization of the code is delicate and has been pioneered in Ref.~\cite{White89}  for the finite-temperature algorithm and in Refs.~\cite{Sugiyama86,Sorella89} for the zero temperature projective algorithm.
As shown in Ref.~\cite{Assaad08_rev}  scales can be omitted in the ground state algorithm -- thus rendering it very stable --  but have to be taken into account in the  finite-temperature code. Apart from runtime information, the file \texttt{info} contains important information concerning the stability of the code.
It is important to know that numerical stabilization is delicate and there is no guarantee  that it will work for all models.

If the numerical stabilization turns out to be bad, one option is to reduce the  value of the parameter \texttt{Nwrap} in the parameter file. 
For performing the stabilization of the involved matrix multiplications we rely on routines from LAPACK. Hence it is very likely that your results may change significantly if you switch the LAPACK implementation.
In order to offer a simple baseline to which people can quickly switch if they want to see whether their results depend on the library used for linear algebra routines we have included parts of the LAPACK-3.6.1 reference implementation from
\url{http://www.netlib.org/lapack/}. You can switch to the QR decomposition related routines from the LAPACK reference implementation by including the switch \texttt{-DQRREF} into the PROGRAMCONFIGURATION string.
To use these routines you need to link against a lapack library that implements at least the LAPACK-3.4.0 interface.\footnote{ We have encountered some compiling issues with this flag. In particular  the  older  intel  ifort  compiler version 10.1  fails for all optimization levels.}

To provide further flexibility, we have  kept the history of different stabilization schemes.   Our default strategy is quick and generically works well but we have  encountered some  models where  it  fails.   If this applies to your model, you can use the switch 
\texttt{-DSTAB2} (stabilization scheme based on the QR decomposition, but not using the LAPACK reference implementation) or  
\texttt{-DSTAB1} (stabilization scheme based on singular value decomposition)   in the header of the file \texttt{Makefile} and recompile the code.  

Typical values for the numerical precision can be found in the examples of Sec.~\ref{sec:ex} (see Sec.~\ref{sec:prec_charge} and \ref{sec:prec_spin}).
%
\subsection{Scripts}\label{sec:scripts}
%
\begin{table}[h]
   \begin{tabular}{@{} l l l @{}}\toprule
   Script & Description & Section\\\midrule
   \path{setenv.sh} & Exports the path variable. &  \ref{sec:running} \\
   \path{Start/analysis.sh} & Starts the error analysis. & \ref{sec:analysis}, \ref{sec:running} \\
   \path{Start/out_to_in.sh} & Copies the output configurations of HS and Ising spins &\\
   & to the respective input files. & \ref{sec:running} \\
   \bottomrule
   \end{tabular}
   \caption{Overview of the bash script files. 
      \label{table:scripts}}
\end{table}

%
\subsection{Analysis programs }\label{sec:analysis}
%
\begin{table}[h]
  \begin{tabular}{@{} l l @{}}\toprule
   Program & Description \\\midrule
   \texttt{cov\_scal.f90}  &  In combination with the script \texttt{analysis.sh}, the bin files with suffix \texttt{\_scal}\\
                           & are read in, and  the corresponding files with suffix \texttt{\_scalJ} are produced.\\
                           & They  contain the  result of the Jackknife rebinning analysis  (see Sec.~\ref{sec:sampling}).  \\
   \texttt{cov\_eq.f90}    &  In combination with the script \texttt{analysis.sh}, the bin files with suffix \texttt{\_eq}\\
                           & are read in, and the corresponding files with suffix  \texttt{\_eqJR}  and  \texttt{\_eqJK}  are\\
                           & produced. They  correspond  to correlation functions in real and Fourier\\
                           & space, respectively.  \\
   \texttt{cov\_tau.f90}   &  In combination with the script \texttt{analysis.sh}, the bin files  \texttt{X\_tau} are read in, \\
                           & and the directories  \texttt{X\_kx\_ky} are produced  for all \texttt{kx} and \texttt{ky} greater or equal\\
                           & to zero. Here \texttt{X}  is a place holder from \texttt{Green}, \texttt{SpinXY}, etc   as specified in\\
                           & \texttt{Alloc\_obs(Ltau)} (See section \ref{Alloc_obs_sec}). Each directory contains  a  file\\
                           & \texttt{g\_kx\_ky}  containing the  time displaced correlation function traced over the\\
                           & orbitals. It also contains the   covariance matrix if \texttt{N\_cov} is set to unity in\\
                           & the parameter file (see Sec.~\ref{sec:input}). \\
                           & Equally, a directory  \texttt{X\_R0}  for the local  time displaced  correlation function\\
                           & is generated. \\\bottomrule
   \end{tabular}
   \caption{ Overview of analysis programs that are called within the script \texttt{analysis.sh}. \label{table:analysis_programs}}
\end{table}
Here we briefly   discuss the analysis programs which read in bin/s and carry out the error analysis. (See Sec.~\ref{sec:sampling}  for a more detailed discussion.)
Error analysis   is based  on the central limit theorem,  which requires bins to be statistically independent, and also the existence of a well-defined variance  for the observable under consideration. 
The former will be the case if bins are  longer than the autocorrelation time.  The latter has to be checked by the user.  In the parameter file listed in Sec.~\ref{sec:input}, the user  can specify how many initial bins should be omitted (variable \texttt{n\_skip}). 
This  number should be at least comparable or  lager than the autocorrelation time. The analysis of the autocorrelation time is triggered by specifying a positive value for \path{N_auto} that is turned off be default ($\textrm{\path{N_auto}}=0$).
The  rebinning  variable \texttt{N\_rebin} will merge \texttt{N\_rebin}  bins into a single new bin. 
If the autocorrelation time  is smaller than the effective bin size, the error should become independent of the bin size and thereby of the variable \texttt{N\_rebin}.  
Our analysis is based on the Jackknife resampling\cite{efron1981,Assaad02}, which includes proper treatment of the sign.
As listed in Table  \ref{table:analysis_programs}  we provide three analysis programs to account for the three observable types. The programs can be found in the directory \texttt{Analysis}  and   are executed by running the  bash shell script 
\texttt{analysis.sh}.
\begin{table}[h]
   \begin{tabular}{@{} l l @{}}\toprule
   File & Description \\\midrule
   \texttt{parameters}  &  Contains also variables for the error analysis:\\
   & \texttt{n\_skip}, \texttt{N\_rebin}, \texttt{N\_Cov} and \texttt{N\_auto} (see Sec.~\ref{sec:input}) \\
   \texttt{X\_scal}, \texttt{Y\_eq}, \texttt{Y\_tau} & Monte Carlo bins (see Table \ref{table:output}) \\\bottomrule
    \end{tabular}
   \caption{Standard input files for the error analysis. \label{table:analysis_input}}
\end{table}
\begin{table}[h]
   \begin{tabular}{@{} l l l @{}}\toprule
   File & Description \\\midrule
   \texttt{X\_scalJ} & Jackknife mean and error of \texttt{X}, where  \texttt{X} stands for \texttt{Kin, Pot, Part},\\ & and \texttt{Ener}.\\
   \texttt{X\_scal\_Auto\_N} & QMC-time resolved autocorrelation and rebinning analysis of \texttt{X},\\ & where  \texttt{X} stands for \texttt{Kin, Pot, Part}, and \texttt{Ener} and \texttt{N} labels the\\ & 
   component if \texttt{X} is a vector.\\
   \texttt{Y\_eqJR} and \texttt{Y\_eqJK} & Jackknife mean and error of \texttt{Y}, where \texttt{Y} stands for \texttt{Green, SpinZ,}\\ & \path{SpinXY}, and \texttt{Den}.\\
   & The suffixes \texttt{R} and \texttt{K} refer to real and reciprocal space, respectively.\\
   \texttt{Y\_R0/g\_R0} & Time-resolved and spatially local Jackknife mean and error of \texttt{Y},\\
   & where \texttt{Y} stands for \texttt{Green, SpinZ, SpinXY}, and \texttt{Den}.\\
   \texttt{Y\_eq\_Auto\_Tr\_kx\_ky} & QMC-time resolved autocorrelation and rebinning analysis of \texttt{Y},\\
   & where \texttt{Y} stands for \texttt{Green, SpinZ, SpinXY}, and \texttt{Den}.\\
   \texttt{Y\_kx\_ky/g\_kx\_ky} & Time resolved and $\vec{k}$-dependent Jackknife mean and error of \texttt{Y},\\
   & where \texttt{Y} stands for \texttt{Green, SpinZ, SpinXY}, and \texttt{Den}.\\\bottomrule
    \end{tabular}
   \caption{ Standard output files of the error analysis. \label{table:analysis_output}}
\end{table}
In the following, we describe the formatting of the output files mentioned in Table \ref{table:analysis_output}.
\begin{itemize}
\item For the scalar quantities \texttt{X}, the output files  \texttt{X\_scalJ} have the following formatting:
\begin{alltt}
Effective number of bins, and bins:      <N_bin - n_skip>      <N_bin>

OBS :    1      <mean(X)>      <error(X)>

OBS :    2      <mean(sign)>   <error(sign)>
\end{alltt}

\item For the autocorrelation analysis of scalar quantities \texttt{X}, the output files  \texttt{X\_scal\_Auto\_N} have the following formatting:
\begin{alltt}
	do i = 1, N_auto
	   tau(i)/n_rebin   Auto_X(tau)   <error( X )>
	enddo
\end{alltt}

\item For the equal time correlation functions \texttt{Y}, the formatting of the output files \texttt{Y\_eqJR} and \texttt{Y\_eqJK} follows this structure:
\begin{alltt}
do i = 1, N_unit_cell
   <k_x(i)>   <k_y(i)>
   do alpha = 1, N_orbital
   do beta  = 1, N_orbital
      alpha  beta  Re<mean(Y)>  Re<error(Y)>  Im<mean(Y)>  Im<error(Y)>
   enddo
   enddo
enddo
\end{alltt}
where \texttt{Re} and \texttt{Im} refer to the real and imaginary part, respectively.

\item For the autocorrelation analysis of equal time quantities \texttt{Y}, the output files  \texttt{Y\_eq\_Auto\_Tr\_kx\_ky} have the following formatting:
\begin{alltt}
	do i = 1, N_auto
	   tau(i)/n_rebin   Auto_Tr[Y](tau)   <error( Tr[Y] )>
	enddo
\end{alltt}

\item The imaginary-time displaced correlation functions \texttt{Y} are written to the output files \texttt{Y\_R0/g\_R0}, when measured locally in space, 
and to the output files \texttt{Y\_kx\_ky/g\_kx\_ky} when they are measured $\vec{k}$-resolved. 
Both output files have the following formatting:
\begin{alltt}
do i = 0, Ltau
   tau(i)   <mean( Tr[Y] )>   <error( Tr[Y])>
enddo
\end{alltt}
where \texttt{Tr} corresponds to the trace over the orbital degrees of freedom.

\end{itemize}

%
\subsection{Running the code}\label{sec:running}
%
In this section we describe the steps how to compile and run the code, as well as how to perform the error analysis of the data.
%
\subsubsection{Compilation}
%
The environment variables and the directives to compile the code are set in the following makefile \texttt{Makefile}:
\lstset{style=bash}
\begin{lstlisting}

# -DMPI selects MPI.
# -DSTAB1   Alternative stabilization, using the singular value decomposition.
# -DSTAB2   Alternative stabilization, lapack QR with  manual pivoting.
#           Packed form of QR factorization is not used.
# (no flag) Default  stabilization, using lapack QR with pivoting. 
#           Packed form of QR factorization  is used. 
# -DQRREF   Enables reference lapack implementation of QR decomposition.
# Recommendation: just use the -DMPI flag if you want to run in parallel or 
#                 leave it empty for serial jobs.  
#                 The default stabilization, no flag, is generically the best. 
PROGRAMCONFIGURATION = -DMPI 
PROGRAMCONFIGURATION = 
f90 = gfortran
export f90
F90OPTFLAGS = -O3 -Wconversion  -fcheck=all
F90OPTFLAGS = -O3
export F90OPTFLAGS
F90USEFULFLAGS = -cpp -std=f2003
F90USEFULFLAGS = -cpp
export F90USEFULFLAGS
FL = -c ${F90OPTFLAGS} ${PROGRAMCONFIGURATION}
export FL
DIR = ${CURDIR}
export DIR
Libs = ${DIR}/Libraries/
export Libs
LIB_BLAS_LAPACK = -llapack -lblas
export LIB_BLAS_LAPACK

all: lib ana program

lib:
	cd Libraries && $(MAKE)
ana:
	cd Analysis && $(MAKE)
program:
	cd Prog && $(MAKE)


clean: cleanall
cleanall: cleanprog cleanlib cleanana
cleanprog:
	cd Prog && $(MAKE) clean
cleanlib:
	cd Libraries && $(MAKE) clean
cleanana:
	cd Analysis && $(MAKE) clean
help:
	@echo "The following are some of the valid targets of this Makefile"
	@echo "all, program, lib, ana, clean, cleanall, cleanprog, cleanlib,
	       cleanana"

\end{lstlisting}
In the above, the GNU Fortan compiler \texttt{gfortran} is set.\footnote{A known issue with the alternative Intel Fortran compiler \texttt{ifort} is the handling of automatic, temporary arrays 
which \texttt{ifort} allocates on the stack. For large system sizes and/or low temperatures this may lead to 
a runtime error. One solution is to demand allocation of arrays above a certain size on the heap instead of the stack. 
This is accomplished by the \texttt{ifort} compiler flag \texttt{-heap-arrays [n]} where \texttt{[n]} is the minimal size (in kilobytes, for example \texttt{n=1024}) of arrays 
that are allocated on the heap.}
We provide a set of options for compilation of the QMC code. The present options are \texttt{-DMPI}, \texttt{-DQRREF}, \texttt{-DSTAB1}, and \texttt{-DSTAB2}. 
They can be included in the string variable \texttt{PROGRAMCONFIGURATION} by the user, as shown above.
The program can be compiled and ran either in single-thread mode (default) or 
in multi-threading mode (define \texttt{-DMPI}) using the MPI standard for parallelization. The remaining three compiler options select a particular stabilization scheme for the matrix multiplications (see Sec.~\ref{sec:output_prec}).
To compile the libraries, the analysis routines and the QMC program at once, just execute the single command:
\begin{verbatim}
make
\end{verbatim}
To clean up all directories and remove the object files and executables, execute the command \texttt{make clean}. As can be seen in the above makefile, there exist also rules to compile/clean up the library, the analysis routines and the QMC program separately.  

%
\subsubsection{Starting a simulation}
%
To start a simulation from scratch, the following files have to be present: \texttt{parameters} and \texttt{seeds}. 
To run a single-thread simulation, for example by using the parameters of one of the  Hubbard models described in Sec.~\ref{sec:ex}, issue the command
\begin{verbatim}
./Prog/Examples.out
\end{verbatim}
To restart the code using an existing simulation as a starting point, first run the script \texttt{out\_to\_in.sh} to set 
the input configuration files.
%
\subsubsection{Error analysis}
%
Note that the error analysis script requires the presence of the environment variable \path{DIR} which defines the path to the error analysis programs.
So before starting the error analysis, one has to make this variable available which is done by the script \path{setenv.sh}. The command is
\begin{verbatim}
source ./setenv.sh
\end{verbatim}
To perform an error analysis based on the Jackknife resampling method (Sec.~\ref{sec:jack})  of the Monte Carlo bins for all observables run the script \texttt{analysis.sh} 
(see Sec.~\ref{sec:analysis}). In case that the parameter \path{N_auto} is set to a finite value the script will also trigger the computation of autocorrelation functions (Sec.~\ref{sec:autocorr}).

\section{Examples}\label{sec:ex}
%
\subsection{The $SU(2)$-Hubbard model on a square lattice}\label{sec:walk1}
%
To implement a Hamiltonian, the user has to provide  a module   which  specifies the lattice, the model, as well as the observables  they wish to compute. 
In this section, we describe the module 
\path{Hamiltonian_Examples.f90} which contains an implementation of the Hubbard model on the square lattice. 
A sample run for this model can be found in \path{Examples/Hubbard_SU2_Square/}.
The input files are \path{parameters} and \path{seeds} (see Tab.~\ref{table:input}). The output files are \path{info}, \path{confout}, and files with suffixes \path{_scal}, \path{_eq}, and \path{_tau} that 
contain the raw measurements (see Tab.~\ref{table:output}).

The Hamiltonian reads 
\begin{equation}
\label{eqn_hubbard_sun}
\mathcal{H}= 
\sum\limits_{\sigma=1}^{2} 
\sum\limits_{x,y =1 }^{N_{\text{unit cell}}} 
  c^{\dagger}_{x \sigma} T_{x,y}c^{\phantom\dagger}_{y \sigma} 
+ \frac{U}{2}\sum\limits_{x}\left[
\sum\limits_{\sigma=1}^{2}
\left(  c^{\dagger}_{x \sigma} c^{\phantom\dagger}_{x \sigma}  -1/2 \right) \right]^{2}\;.
\end{equation} 
We can make contact with the general form of the Hamiltonian by setting: 
$N_{\mathrm{fl}} = 1$, $N_{\mathrm{col}} \equiv \texttt{N\_SUN}     =2 $,   $M_T    =    1$,  $T^{(ks)}_{x y}   =  T_{x,y}$,  $M_V   =  N_{\text{unit cell}} $,  $U_{k}       =   -\frac{U}{2}$, 
 $V_{x y}^{(ks)} =  \delta_{x,y} \delta_{x,k}$,  $\alpha_{ks}   = - \frac{1}{2}  $ and $M_I       = 0 $.
%
\subsubsection{Setting the Hamiltonian:  \texttt{Ham\_set} }
%
The main program will call the subroutine   \texttt{Ham\_set} in the module \texttt{Hamiltonian\_Hub.f90}.
The latter  subroutine  defines the  public variables
\lstset{style=fortran}
\begin{lstlisting}

Type (Operator), dimension(:,:), allocatable  :: Op_V 
Type (Operator), dimension(:,:), allocatable  :: Op_T
Integer, allocatable :: nsigma(:,:)
Integer              :: Ndim,  N_FL,  N_SUN,  Ltrot

\end{lstlisting}
which specify the model. The array \texttt{nsigma} contains the HS field. The  routine \texttt{Ham\_set}  will first  read the parameter file,  then set the lattice, \texttt{Call Ham\_latt},  set the hopping, \texttt{Call Ham\_hop},  and set the interaction, 
\texttt{call Ham\_V}.  
The parameters are read in from the file \texttt{parameters}, see Sec.~\ref{sec:input}.
%
\paragraph{The lattice:   \texttt{Call Ham\_latt} }
%
The choice \texttt{Lattice\_type = "Square"} sets $\vec{a}_1 =  (1,0) $ and $\vec{a}_2 =  (0,1) $  and for an $L_1 \times L_2$  lattice  $\vec{L}_1 = L_1 \vec{a}_1$ and  $\vec{L}_2 = L_2 \vec{a}_2$.     The call to  \texttt{Call Make\_Lattice( L1, L2, a1,  a2, Latt)} will generate the lattice   \texttt{Latt} of type \texttt{Lattice}. 
For the Hubbard model on the square lattice, the number of orbitals per unit cell is given by \texttt{NORB=1} such that   $N_{\mathrm{dim}}   \equiv N_{\text{unit cell}}   \cdot \texttt{NORB}  = \texttt{Latt\%N} \cdot \texttt{NORB}$, since $N_{\text{unit cell}} = \texttt{Latt\%N}$.

%
\paragraph{The hopping term: \texttt{Call Ham\_hop}}
%
The hopping matrix is implemented as follows. 
We allocate an array of dimension $1\times 1$ of type operator  called \texttt{Op\_T} and set the  dimension for the hopping  matrix to $N=N_{\mathrm{dim}}$. One  allocates and initializes this type by a single call to the subroutine \texttt{Op\_make}: 
\begin{lstlisting}

call Op_make(Op_T(1,1),Ndim)

\end{lstlisting}
Since the hopping  does not  break down into small blocks, we have ${\bm P}=\mathds{1}$   and  
\begin{lstlisting}

Do i= 1,Ndim
  Op_T(1,1)%P(i) = i
Enddo

\end{lstlisting}
We set the hopping matrix  with 

\begin{lstlisting}

DO I = 1, Latt%N
   Ix = Latt%nnlist(I,1,0)
   Iy = Latt%nnlist(I,0,1)
   Op_T(1,1)%O(I  ,Ix) = cmplx(-Ham_T,   0.d0,kind(0.D0))
   Op_T(1,1)%O(Ix, I ) = cmplx(-Ham_T,   0.d0,kind(0.D0))
   Op_T(1,1)%O(I  ,Iy) = cmplx(-Ham_T,   0.d0,kind(0.D0))
   Op_T(1,1)%O(Iy, I ) = cmplx(-Ham_T,   0.d0,kind(0.D0))
   Op_T(1,1)%O(I  ,I ) = cmplx(-Ham_chem,0.d0,kind(0.D0))
ENDDO

\end{lstlisting}
Here, the integer function \texttt{  j=  Latt\%nnlist(I,n,m)} is defined in the lattice module and returns the index of the lattice site $ \vec{I} +  n \vec{a}_1 +  m \vec{a}_2$.
Note that periodic boundary conditions are 
already taken into account.  The hopping parameter \texttt{Ham\_T} as well as the chemical potential \texttt{Ham\_chem} are read from the parameter file.  
To completely define the hopping  we further set: \texttt{Op\_T(1,1)\%g = -Dtau }, \texttt{Op\_T(1,1)\%alpha = cmplx(0.d0,0.d0, kind(0.D0))} and call the routine  \texttt{Op\_set(Op\_T(1,1))}  so as to generate  the unitary transformation and eigenvalues as specified in Table \ref{table:operator}.  Recall that for the hopping, the variable  \texttt{Op\_set(Op\_T(1,1))\%type}  is not  required. 
Note that although a checkerboard decomposition is not  used here,  it can be implemented by considering a larger number of sparse hopping matrices.

%
\paragraph{The interaction term: \texttt{Call Ham\_V}}
%
To implement this interaction, we allocate an array of \texttt{Operator} type. The array is called  \texttt{Op\_V} and has dimensions $N_{\mathrm{dim}}\times N_{\mathrm{fl}}=N_{\mathrm{dim}} \times 1$. 
We set the dimension for the interaction term to  $N=1$, and  allocate and initialize this array of type  \texttt{Operator} by repeatedly calling the subroutine \texttt{Op\_make}: 

\begin{lstlisting}

do i  = 1,Ndim
   call Op_make(Op_V(i,1),1)
enddo

\end{lstlisting}
For each lattice site $i$, the  matrices ${\bm P}$ are of dimension $1\times N_{\mathrm{dim}} $ and have only one non-vanishing entry. Thereby we can set:

\begin{lstlisting}

Do i = 1,Ndim
   Op_V(i,1)%P(1)   = i
   Op_V(i,1)%O(1,1) = cmplx(1.d0,0.d0, kind(0.D0))
   Op_V(i,1)%g      = sqrt(cmplx(-dtau*ham_U/dble(N_SUN),0.D0,kind(0.D0)))
   Op_V(i,1)%alpha  = cmplx(-0.5d0,0.d0, kind(0.D0))
   Op_V(i,1)%type   = 2
   Call Op_set( Op_V(i,1) )
Enddo

\end{lstlisting}
so as to completely define the interaction term. 

%
\subsubsection{Observables}
%
At this point, all the information   for the simulation to  start has been provided.  The code will sequentially go through  the operator list  \texttt{Op\_V}  and update the  fields.   Between  time slices   \texttt{LOBS\_ST}  and  \texttt{LOBS\_EN}   the main program will call the routine  \texttt{Obser(GR,Phase,Ntau)}   which is provided by the user and handles equal time correlation functions. 
If \texttt{Ltau=1} the main program will call the routine \texttt{ObserT(NT,  GT0,G0T,G00,GTT, PHASE) }   which is again  provided by the user and handles  imaginary-time displaced correlation functions. 

The user will have to  implement the  observables  he/she  wants to compute. Here  we  will describe how to  proceed. 
%
\paragraph{Allocating space for the observables: \texttt{Call Alloc\_obs(Ltau) }} \label{Alloc_obs_sec}
%
For  four scalar  or vector observables,  the user will have to  declare the following: 
\begin{lstlisting}

Allocate ( Obs_scal(4) )
Do I = 1,Size(Obs_scal,1)
   select case (I)
   case (1)
      N = 2;  Filename ="Kin"
   case (2)
      N = 1;  Filename ="Pot"
   case (3)
      N = 1;  Filename ="Part"
   case (4)
      N = 1,  Filename ="Ener"
   case default
      Write(6,*) ' Error in Alloc_obs '  
   end select
   Call Obser_Vec_make(Obs_scal(I),N,Filename)
enddo
\end{lstlisting}
Here,   \texttt{Obs\_scal(1)}   contains a vector  of two observables  so as to account for the $x$- and $y$-components of the kinetic energy for example.  

For equal time correlation  functions  we allocate  \texttt{Obs\_eq}  of type \texttt{Obser\_Latt}.  Here we include the calculation of spin-spin and density-density correlation functions alongside equal time Green functions. 
\begin{lstlisting}

Allocate ( Obs_eq(4) )
Do I = 1,Size(Obs_eq,1)
   select case (I)
   case (1)
      Ns = Latt%N; No = Norb;  Filename ="Green"
   case (2)
      Ns = Latt%N; No = Norb;  Filename ="SpinZ"
   case (3)
      Ns = Latt%N; No = Norb;  Filename ="SpinXY"
   case (4)
      Ns = Latt%N; No = Norb;  Filename ="Den"
   case default
      Write(6,*) ' Error in Alloc_obs '  
   end select
   Nt = 1
   Call Obser_Latt_make(Obs_eq(I),Ns,Nt,No,Filename)
enddo
 \end{lstlisting} 
 For the Hubbard model \texttt{Norb = 1} and for   equal time correlation functions   \texttt{Nt = 1}.       If  \texttt{Ltau = 1}  then the code will allocate space for  time displaced quantities.   The same structure as for  equal time correlation functions will be used albeit with  \texttt{Nt = Ltrot + 1}.  At the beginning of each bin, the main program will set the bin observables to zero by calling  the routine 
 \texttt{Init\_obs(Ltau)}.   The user does not have to edit this routine. 
%
\paragraph{Measuring equal time observables: \texttt{Obser(GR,Phase,Ntau)}}
%
The equal time  Green function,
\begin{equation}
	 \texttt{GR(x,y},\sigma{\texttt)}  = \langle c^{\phantom{\dagger}}_{x,\sigma} c^{\dagger}_{y,\sigma}  \rangle,
\end{equation}
the  phase factor \texttt{phase} [Eq.~(\ref{eqn:phase})], and time slice \texttt{Ntau}   are provided by the main program.  
Here, $x$ and $y$ label  both unit cell as well as the orbital within the unit cell. For the Hubbard model described here, $x$ corresponds to the unit cell.  The Green function  does not depend on the color index, and is diagonal in flavor.  For the $SU(2)$ symmetric implementation  there is only one flavor, $\sigma = 1$ and the Green function is  independent on the spin index.  This renders the calculation of the observables particularly easy.   

An explicit calculation of the   potential energy  $ \langle U \sum_{\vec{i}}  \hat{n}_{\vec{i},\uparrow}   \hat{n}_{\vec{i},\downarrow}  \rangle $ reads 

\begin{lstlisting} 

Obs_scal(2)%N        = Obs_scal(2)%N + 1
Obs_scal(2)%Ave_sign = Obs_scal(2)%Ave_sign + Real(ZS,kind(0.d0))
Do i = 1,Ndim
   Obs_scal(2)%Obs_vec(1)=Obs_scal(2)%Obs_vec(1)+(1-GR(i,i,1))**2*Ham_U*ZS*ZP
Enddo

\end{lstlisting} 
Here  $ \texttt{ZS} = \text{ sign} (C) $  [see Eq.~(\ref{Sign.eq})],  $ \texttt{ZP} =   \frac{e^{-S(C)}} {\Re \left[e^{-S(C)} \right]}   $ [see Eq.~(\ref{eqn:phase})] and  \texttt{Ham\_U}  corresponds to the Hubbard-$U$ term.\\
Equal time correlations  are also computed in this routine. As an explicit example, we  consider the equal time density-density correlation:
\begin{equation}
	 \langle n_{\vec{i},\alpha}   n_{\vec{j},\beta} \rangle   -  \langle n_{\vec{i},\alpha} \rangle  \langle    n_{\vec{j},\beta}  \rangle \;.
\end{equation} 
For the calculation of such quantities, it is convenient to  define: 
\begin{equation}
\label{GRC.eq}
	\texttt{GRC(x,y,s)}   =  \delta_{x,y}  - \texttt{GR(y,x,s)  }
\end{equation}
such that \texttt{GRC(x,y,s)}    corresponds to  $ \langle \langle  \hat{c}_{x,s}^{\dagger}\hat{c}_{y,s}^{\phantom\dagger} \rangle \rangle $. 
In the program code, the calculation of the equal time density-density correlation function looks as follows:
\begin{lstlisting} 

Obs_eq(4)%N     = Obs_eq(4)%N + 1  ! Even if it is redundant, each observable  
                                   ! carries its own counter and sign.
Obs_eq(4)%Ave_sign = Obs_eq(4)%Ave_sign + Real(ZS,kind(0.d0))  
Do I1 = 1,Ndim
   I    = List(I1,1)               ! = I1  (The Hubbard model  on the square
   no_I = List(I1,2)               ! = 1   lattice has one orbital per unit
                                   !       cell)
   Do J1 = 1,Ndim                       
      J    = List(J1,1)
      no_J = List(J1,2)
      imj = latt%imj(I,J)
      Obs_eq(4)%Obs_Latt(imj,1,no_I,no_J) = &
              & Obs_eq(4)%Obs_Latt(imj,1,no_I,no_J) + &
              &     (    GRC(I1,I1,1) * GRC(J1,J1,1) * N_SUN * N_SUN      + &
              &          GRC(I1,J1,1) * GR(I1,J1,1) * N_SUN   ) * ZP * ZS 
   Enddo
   Obs_eq(4)%Obs_Latt0(no_I) = &
   & Obs_eq(4)%Obs_Latt0(no_I)+GRC(I1,I1,1) * N_SUN * ZP * ZS
Enddo
\end{lstlisting} 
Note that we consider the  square lattice of the single site Hubbard model as a special case of a multiorbital problem as described in Sec.~\ref{sec:multi-orbital}
At the end of each bin  the main program will call the routine \texttt{ Pr\_obs(LTAU)}. This routine will append the result of the bins in the specified file,  with appropriate suffix. 
%
\paragraph{Measuring time displaced observables: \texttt{ObserT(NT,  GT0,G0T,G00,GTT, PHASE) }}
%
This subroutine is called by the main program at the beginning of each sweep, provided that \texttt{LTAU}  is set to unity.  \texttt{NT} runs from \texttt{0}  to \texttt{Ltrot} and denotes the   imaginary time difference.   For a given time  displacement, the main program provides:
\begin{eqnarray}
\texttt{GT0(x,y,s) }  &=&   \phantom{+} \langle \langle \hat{c}^{\phantom\dagger}_{x,s} (Nt \Delta \tau)   \hat{c}^{\dagger}_{y,s} (0)   \rangle \rangle = \langle \langle {\cal T} \hat{c}^{\phantom\dagger}_{x,s} (Nt \Delta \tau)   \hat{c}^{\dagger}_{y,s} (0)   \rangle \rangle  \nonumber \\
\texttt{G0T(x,y,s) }   &=&  -   \langle \langle   \hat{c}^{\dagger}_{y,s} (Nt \Delta \tau)    \hat{c}^{\phantom\dagger}_{x,s} (0)    \rangle \rangle =
    \langle \langle {\cal T} \hat{c}^{\phantom\dagger}_{x,s} (0)    \hat{c}^{\dagger}_{y,s} (Nt \Delta \tau)   \rangle \rangle  \nonumber  \\
  \texttt{G00(x,y,s) }  &=&    \phantom{+} \langle \langle \hat{c}^{\phantom\dagger}_{x,s} (0)   \hat{c}^{\dagger}_{y,s} (0)   \rangle \rangle    \nonumber \\
    \texttt{GTT(x,y,s) }  &=&   \phantom{+} \langle \langle \hat{c}^{\phantom\dagger}_{x,s} (Nt \Delta \tau)   \hat{c}^{\dagger}_{y,s} (Nt \Delta \tau)   \rangle \rangle    
\end{eqnarray}
In the above we have omitted the color index since  the  Green functions are color independent.  The time displaced  spin-spin correlations 
$ 4 \langle \langle \hat{S}^{z}_{\vec{i}} (\tau)  \hat{S}^{z}_{\vec{j}} (0)\rangle \rangle   $ 
are thereby given by: 
\begin{equation}
	4 \langle \langle \hat{S}^{z}_{\vec{i}} (\tau)  \hat{S}^{z}_{\vec{j}} (0)\rangle \rangle   = - 2 \; \texttt{G0T(J1,I1,1) } \texttt{GT0(I1,J1,1) } \;.
\end{equation}
Note that the above holds for the $SU(2)$ HS transformation discussed in this chapter. 
The handling of time displaced correlation functions is identical to that of equal time correlations. 
%
\subsubsection{Numerical precision}\label{sec:prec_charge}
%
The directory \path{Examples/Hubbard_SU2_Square}  contains an example simulation of the $4 \times 4$ Hubbard model at $U/t=4$ and $\beta t = 10$. 
Information on the numerical stability is included in the following lines of the corresponding file \texttt{info}:
\begin{alltt}
Precision Green  Mean, Max : 1.2918865817224671E-014  4.0983018995027644E-011
Precision Phase, Max       : 5.0272908791449966E-012
Precision tau    Mean, Max : 8.4596701790588625E-015  3.5033530012121281E-011
\end{alltt}
showing the mean and maximum difference between the \textit{wrapped}  and from scratched computed equal and time displaced  Green functions \cite{Assaad08_rev}.
A stable code  should produce results where the mean difference is smaller than the  stochastic error. The above example  shows a very stable  simulation since the Green function  is of order one. 

%
\subsection{The $M_z$-Hubbard model on a square lattice}\label{sec:walk1.1}
%
The Hubbard Hamiltonian can equally be written as:
\begin{equation}
\label{eqn_hubbard_Mz}
\mathcal{H}=
\sum\limits_{\sigma=1}^{2} 
\sum\limits_{x,y =1 }^{N_{unit\; cells }} 
  c^{\dagger}_{x \sigma} T_{x,y}c^{\phantom\dagger}_{y \sigma} 
- \frac{U}{2}\sum\limits_{x}\left[
c^{\dagger}_{x, \uparrow} c^{\phantom\dagger}_{x \uparrow}  -   c^{\dagger}_{x, \downarrow} c^{\phantom\dagger}_{x \downarrow}  \right]^{2}\;.
\end{equation} 
We can make contact with the general form of the Hamiltonian  (see Eq.~\ref{eqn:general_ham}) by setting: 
$N_{\mathrm{fl}} = 2$, $N_{\mathrm{col}} \equiv \texttt{N\_SUN}     =1 $,   $M_T    =    1$,  $T^{(ks)}_{x y}   =  T_{x,y}$,  $M_V   =  N_{\text{unit cell}} $,  $U_{k}       =   \frac{U}{2}$, 
 $V_{x y}^{(k, s=1)} =  \delta_{x,y} \delta_{x,k}  $,  $V_{x y}^{(k, s=2)} =  - \delta_{x,y} \delta_{x,k}  $,  $\alpha_{ks}   = 0  $ and $M_I       = 0 $.   
 The coupling of the HS fields to the $z$-component of   the magnetization breaks the $SU(2)$ spin symmetry. Nevertheless the $z$-component of the spin remains a good quantum number such that the imaginary-time propagator -- for a given HS field -- is block  diagonal in this quantum number. This corresponds to the flavor index  which runs from one to two  labelling spin up and spin down  degrees of freedom.       In the parameter file  listed in  Sec.~\ref{sec:input}  setting the model variable to  \texttt{Hubbard\_Mz}  will carry out the simulation in the above representation. 
 With respect to the $SU(2)$ case, the changes required in the \texttt{Hamiltonian\_Examples.f90}   module are  minimal and essentially effect only the interaction term, and the calculation of observables.  We note that  in this formulation the  hopping matrix can be flavor dependent such that a Zeeman  magnetic field can be introduced.  If the chemical potential is set to zero, this will not generate a negative sign problem \cite{Wu04,Milat04,Bercx09}.    
 A sample run for this model can be found in \texttt{Examples/Hubbard\_Mz\_Square/}.
The input files are \path{parameters} and \path{seeds} (see Tab.~\ref{table:input}). The output files are \path{info}, \path{confout}, and files with suffixes \path{_scal}, \path{_eq}, and \path{_tau} that 
contain the raw measurements (see Tab.~\ref{table:output}).
%
\subsubsection{The interaction term: \texttt{Call Ham\_V} } 
%
The interaction term is now given by: 

\begin{lstlisting}

Allocate(Op_V(Ndim,N_FL))
do nf = 1,N_FL
   do i  = 1, Ndim
      Call Op_make(Op_V(i,nf),1)
   enddo
enddo
Do nf = 1,N_FL
   nc = 0
   X = 1.d0
   if (nf == 2) X = -1.d0
   Do i = 1,Ndim
      nc = nc + 1
      Op_V(nc,nf)%P(1) = I
      Op_V(nc,nf)%O(1,1) = cmplx(1.d0, 0.d0, kind(0.D0))
      Op_V(nc,nf)%g      = X*SQRT(CMPLX(DTAU*ham_U/2.d0, 0.D0, kind(0.D0))) 
      Op_V(nc,nf)%alpha  = cmplx(0.d0, 0.d0, kind(0.D0))
      Op_V(nc,nf)%type   = 2
      Call Op_set( Op_V(nc,nf) )
   Enddo
Enddo

\end{lstlisting}
In the above, one will see explicitly that  there is a sign   difference between  the coupling of the HS field  in  the  two flavor sectors.  
%
 \subsubsection{The  measurements: \texttt{Call Obser, Call  ObserT} } 
%
 Since  the spin up and spin down Green functions differ  for a given HS configuration,  the Wick decomposition will take a different form. In particular, the  equal time spin-spin correlation functions 
 $ 4 \langle \langle \hat{S}^{z}_{\vec{i}}   \hat{S}^{z}_{\vec{j}} \rangle \rangle   $  calculated in the subroutine  \texttt{Obser}  will take the form: 
  \begin{eqnarray}
   4 \langle \langle \hat{S}^{z}_{x}   \hat{S}^{z}_{y} \rangle \rangle   =  & &  \texttt{  GRC(x,y,1) * GR(x,y,1) + GRC(x,y,2) * GR(x,y,2) + }  \nonumber \\ 
& & \texttt{   (GRC(x,x,2) - GRC(x,x,1))*(GRC(y,y,2) - GRC(y,y,1))}  \nonumber
  \end{eqnarray}
Here,  \texttt{GRC}  is defined in Eq.~\ref{GRC.eq}.  Equivalent changes will have to be carried out for other equal time and time displaced observables. 
  
Apart from these modifications, the program  will run in exactly the same manner as for the $SU(2)$ case. 
    
%
\subsubsection{Numerical precision}\label{sec:prec_spin}
%
The directory \path{Examples/Hubbard_Mz_Square}  contains an example simulation of the $4 \times 4$ Hubbard model at $U/t=4$ and $\beta t = 10$. 
Information on the numerical stability is included in the following lines of the corresponding file \texttt{info}:
 \begin{alltt}
Precision Green  Mean, Max : 5.0823874429126405E-011  5.8621144596315844E-006
Precision Phase, Max       : 0.0000000000000000     
Precision tau    Mean, Max : 1.5929357848647394E-011  1.0985132530727526E-005 
\end{alltt}

This is still an excellent precision but nevertheless choosing a 
 HS field which couples to the z-component of the magnetization apparently leads to numerical results that are 
a couple of order of magnitudes less precise than a HS decomposition coupling to the charge (compare with Sec.~\ref{sec:prec_charge}).

%
\subsection{The $SU(2)$-Hubbard model on  the honeycomb  lattice}\label{sec:walk1.2}
%
The Hamilton module \texttt{Hamiltonian\_Examples.f90}   can also carry out simulations for the Hubbard model on the Honeycomb lattice by setting in the parameter file   \path{Lattice_type = "Honeycomb"} (see Sec.~\ref{sec:input}).
 A sample run for this model can be found in \path{Examples/Hubbard_SU2_Honeycomb/}.
The input files are \path{parameters} and \path{seeds} (see Tab.~\ref{table:input}). The output files are \path{info}, \path{confout}, and files with suffixes \path{_scal}, \path{_eq}, and \path{_tau} that 
contain the raw measurements (see Tab.~\ref{table:output}).
%
\subsubsection{Working with multi-orbital unit cells:  \texttt{Call Ham\_Latt} } \label{sec:multi-orbital}
%
  This model is an example of  a multi-orbital unit cell, and the aim of this section is to document how to implement this in the code.  The  Honeycomb lattice is a  triangular Bravais lattice with two orbitals per unit cell.  The routine  \texttt{Ham\_Latt} will set: 

\begin{lstlisting}

Norb    = 2
N_coord = 3
a1_p(1) =  1.D0   ; a1_p(2) =  0.d0
a2_p(1) =  0.5D0  ; a2_p(2) =  sqrt(3.D0)/2.D0             
L1_p    =  dble(L1) * a1_p
L2_p    =  dble(L2) * a2_p
            
\end{lstlisting}
and then  call \path{ Make_Lattice( L1_p, L2_p, a1_p,  a2_p, Latt ) } so as to generate the triangular lattice.  The coordination number of this lattice is \texttt{ N\_coord=3 }  and  the number of orbitals per unit cell  corresponds to \texttt{NORB=2}.    The total number of  orbitals  is thereby: \texttt{$N_{\mathrm{dim}}$=Latt\%N*NORB}.    To easily keep track of the orbital and unit cell, we define a  super-index as shown below:

\begin{lstlisting}

Allocate (List(Ndim,2), Invlist(Latt%N,Norb))
nc = 0
Do I = 1,Latt%N                           ! Unit-cell index 
   Do no = 1,Norb                         ! Orbital index
      nc = nc + 1                         ! Super-index labeling unit cell
                                          ! and orbital
      List(nc,1) = I                      ! Unit-cell  of super index  nc
      List(nc,2) = no                     ! Orbital of super index nc
      Invlist(I,no) = nc                  ! Super-index for given  unit cell
                                          ! and orbital
  Enddo
Enddo

\end{lstlisting}

With the above lists one can run through all the orbitals and at each time keep track of the unit-cell and orbital index.    We note that when translation symmetry is completely absent  one can work with a single unit cell, and the  number of orbitals will then correspond to the  number of lattice sites. 
%
\subsubsection{The hopping term:  \texttt{Call Ham\_Hop} }
%
Some care has to be taken when setting the hopping matrix.    In the Hamilton module \path{Hamiltonian_Examples.f90}   we do this in the following way:

\begin{lstlisting}

DO I = 1, Latt%N                           ! Loop over unit cell 
   do no = 1,Norb                          ! Runs over orbitals and 
                                           ! sets the chemical potential
      I1 = invlist(I,no)  
      Op_T(nc,n)%O(I1 ,I1) = cmplx(-Ham_chem, 0.d0, kind(0.D0))
   enddo
   I1 = Invlist(I,1)                       ! Orbital A of unit cell I
   Do nc1 = 1,N_coord                      ! Loop over coordination  number
      select case (nc1)
      case (1)
         J1 = invlist(I,2)                 ! Orbital B of unit cell i
      case (2)
         J1=invlist(Latt%nnlist(I,1,-1),2) ! Orbital B of unit cell i+a_1-a_2
      case (3)
         J1=invlist(Latt%nnlist(I,0,-1),2) ! Orbital B of unit cell i-a_2 
      case default
         Write(6,*) ' Error in  Ham_Hop '  
      end select
      Op_T(nc,n)%O(I1,J1) = cmplx(-Ham_T,    0.d0, kind(0.D0))   
      Op_T(nc,n)%O(J1,I1) = cmplx(-Ham_T,    0.d0, kind(0.D0))
   Enddo
Enddo

\end{lstlisting}
As apparent from the above, hopping matrix elements   are non-zero only between the $A$ and $B$  sublattices. 
%
\subsubsection{Observables:  \texttt{Call Obser},   \texttt{Call ObserT}}
%
In the multi-orbital case,  the correlation functions have additional orbital indices. This is automatically taken care of in the routines \texttt{Call Obser} and \texttt{Call ObserT}  since  we have already considered the  Hubbard model on the square lattice to correspond to a multi-orbital unit cell albeit with the special choice of one orbital per unit cell. 
%
\subsection{The $SU(2)$-Hubbard model on a square lattice coupled to a transverse Ising field}\label{sec:walk2}
%
The model we consider here  is very similar to the  above,  but has an additional coupling to a transverse field: 
\begin{eqnarray}
\label{eqn_hubbard_sun_Ising}
\mathcal{H}= & & 
\sum\limits_{\sigma=1}^{2} 
\sum\limits_{x,y } 
  c^{\dagger}_{x \sigma} T_{x,y}c^{\phantom\dagger}_{y \sigma} 
+ \frac{U}{2}\sum\limits_{x}\left[
\sum\limits_{\sigma=1}^{2}
\left(  c^{\dagger}_{x \sigma} c^{\phantom\dagger}_{x \sigma}  -1/2 \right) \right]^{2}   
+  \xi \sum_{\sigma,\langle x,y \rangle} \hat{Z}_{\langle x,y \rangle}  \left( c^{\dagger}_{x \sigma} c^{\phantom\dagger}_{y \sigma}  + h.c. \right)  \nonumber \\ 
 & & - h \sum_{\langle x,y \rangle} \hat{X}_{\langle x,y \rangle}   - J \sum_{\langle \langle x,y \rangle \langle x',y' \rangle \rangle} 
  \hat{Z}_{\langle x,y \rangle}   \hat{Z}_{\langle x',y' \rangle} 
\end{eqnarray}
We can make contact with the general form of the Hamiltonian by setting: 
$N_{\mathrm{fl}} = 1$, $N_{\mathrm{col}} \equiv \texttt{N\_SUN}     =2 $,   $M_T    =    1$,  $T^{(ks)}_{x y}   =  T_{x,y}$,  $M_V   =  N_{\text{unit cell}} \equiv N_{\mathrm{dim}}$,  $U_{k}       =   -\frac{U}{2}$, 
 $V_{x y}^{(ks)} =  \delta_{x,y} \delta_{x,k}$,  $\alpha_{ks}   = - \frac{1}{2}  $ and $M_I       = 2 N_{\text{unit cell}} $.  
 The last two terms of the  above Hamiltonian describe a transverse Ising field model on the bonds of the square lattice.  This  type of Hamiltonian  has  recently been extensively discussed  \cite{Schattner15,Xu16,Assaad16}.  Here we adopt the notation of Ref.~\cite{Assaad16}. Note that   $\langle \langle x,y \rangle \langle x',y' \rangle \rangle $ denotes nearest neighbor bonds.
The modifications  required to generalize the Hubbard model code to the above model are two-fold. 
First,  one has to specify the function \path{Real (Kind=8) function S0(n,nt)}, and  second,  modify the interaction \texttt{Call Ham\_V}.
A sample run for this model can be found in \path{Examples/Hubbard_SU2_Ising_Square/}.
The input files are \path{parameters} and \path{seeds} (see Tab.~\ref{table:input}). The output files are \path{info}, \path{confout}, and files with suffixes \path{_scal}, \path{_eq}, and \path{_tau} that 
contain the raw measurements (see Tab.~\ref{table:output}).
%
\subsubsection{The Ising term}
%
Since the Ising field lives on bonds we have to provide a data structure defining this quantity.  A bond has an anchor site as well as an orientation. The routine \path{Setup_Ising_action}   initializes  the arrays \path{L_bond} and  \path{L_bond_inv} that contain this information.

\begin{lstlisting}

nc = 0
Do n_orientation = 1,N_coord
Do I = 1, Latt%N
   nc = nc + 1
   L_bond(I,n_orientation) = nc
   L_bond_inv(nc,1) = I    
   L_bond_inv(nc,2) = n_orientation
enddo
enddo
\end{lstlisting}
The two legs of the bond are given by  the anchor $\vec{I}$ and $\vec{I}+ \vec{a}_{n_\text{orientation}}$.
%
\subsubsection{The interaction term: \texttt{Call Ham\_V}}
%
The dimension of   \texttt{Op\_V}  is now  $(M_I + M_V)\times N_{\mathrm{fl}}=((N_{coord} +  1 )N_{\mathrm{dim}}) \times 1$ since each site has $N_{coord} =2$ bonds   for the square lattice.

\begin{lstlisting}

do i  = 1,N_coord*Ndim                    ! Runs over bonds for Ising inter.
  call Op_make(Op_V(i,1),2)
enddo
do i  =  N_coord*Ndim+1, (N_coord+1)*Ndim ! Runs over sites for Hubbard inter.
  call Op_make(Op_V(i,1),1)
enddo

\end{lstlisting}

The  first  \texttt{N\_coord*Ndim} operators run through the $2N$ bonds of the square lattice and   are given by:
\begin{lstlisting}

Do nc = 1,Ndim*N_coord            ! Runs over bonds. Coordination number = 2.
                                  ! For the square lattice Ndim = Latt%N
  
   I1 = L_bond_inv(nc,1)          ! Anchor of the bond
	                          ! L_bond_inv is setup in Setup_Ising_action
   if ( L_bond_inv(nc,2)  == 1 ) I2 = Latt%nnlist(I1,1,0) ! Second site of
   if ( L_bond_inv(nc,2)  == 2 ) I2 = Latt%nnlist(I1,0,1) ! the bond  
   Op_V(nc,1)%P(1) = I1
   Op_V(nc,1)%P(2) = I2
   Op_V(nc,1)%O(1,2) = cmplx(1.d0 ,0.d0, kind(0.D0))
   Op_V(nc,1)%O(2,1) = cmplx(1.d0 ,0.d0, kind(0.D0))
   Op_V(nc,1)%g      = cmplx(-dtau*Ham_xi,0.D0,kind(0.D0))
   Op_V(nc,1)%alpha  = cmplx(0d0,0.d0, kind(0.D0))
   Op_V(nc,1)%type   = 1
Enddo

\end{lstlisting}
Here,  \texttt{ham\_xi} defines the coupling strength  between the Ising  and fermion degree of freedom.
As for the Hubbard case, the last \texttt{Ndim}  operators read: 

\begin{lstlisting}

nc = N_coord*Ndim 
Do i = 1, Ndim
    nc = nc + 1
    Op_V(nc,1)%P(1)  = i 
    Op_V(nc,1)%O(1,1)= cmplx(1.d0  ,0.d0, kind(0.D0))
    Op_V(nc,1)%g     = sqrt(cmplx(-dtau*ham_U/(DBLE(N_SUN)),0.D0,kind(0.D0)))
    Op_V(nc,1)%alpha = cmplx(-0.5d0,0.d0, kind(0.D0))
    Op_V(nc,1)%type  = 2
Enddo

\end{lstlisting}
%
\subsubsection{The function \texttt{Real (Kind=8) function S0(n,nt)} }\label{sec:s0}
%
As mentioned above,  a configuration now includes both HS spins and Ising spins and is given by
\begin{equation}
	C = \left\{   s_{i,\tau} ,  l_{j,\tau}  \text{ with }  i=1\cdots M_I,  j = 1\cdots M_V,  \tau=1,L_{Trotter}  \right\}\:.
\end{equation}
This configuration is stored in the  integer array \texttt{nsigma(M\_V + M\_I, Ltrot)}.  With the above ordering of Hubbard and Ising interaction terms, and a for a given imaginary time, the first \texttt{2*Ndim} fields correspond to the Ising interaction and the next \path{Ndim} ones to the Hubbard interaction.
The first   argument of the function \texttt{S0}, namely \texttt{n},  corresponds to the index of the operator  string 
\texttt{Op\_V(n,1)}. If \texttt{Op\_V(n,1)\%type = 2} then   \texttt{S0(n,nt)}  returns 1. Note that \texttt{type=2} refers to spins that stem from a  HS transformation. 
If however  \texttt{Op\_V(n,1)\%type = 1}  then function \texttt{S0}  returns
\begin{equation}
\frac{e^{-S_{0,I} \left(  s_{1,\tau},  \cdots,  - s_{n,\tau},  \cdots s_{M_I,\tau}   \right) } }{e^{-S_{0,I}  \left(  s_{1,\tau},  \cdots,   s_{n,\tau},  \cdots s_{M_I,\tau}   \right)   } }	
\end{equation}
That is,   if $n \leq 2* \texttt{Ndim} $,    $ \texttt{S0(n,nt)} $  returns the ratio of the new weight to the old weight  of the  Ising Hamiltonian upon flipping a single Ising spin $ s_{n,\tau} $.  Otherwise, $ \texttt{S0(n,nt)} $   returns unity. 

%
\section{Miscellaneous}\label{sec:misc}
%
\subsection{Other models}\label{sec:other_models}
%
The aim of this section is to briefly mention  a small  selection of  other models that can be studied using the QMC code of the ALF project.  
%
\subsubsection{Kondo lattice model}
%
Simulating the Kondo lattice with the QMC code of the ALF project    requires rewriting of the model along the lines of Refs.~\cite{Assaad99a,Capponi00,Beach04}.  
Adopting the notation of these articles,   the Hamiltonian that one will simulate reads: 
 \begin{equation}\label{eqn:ham_kondo}
 	\hat{\mathcal{H}}  = 
	\underbrace{-t \sum_{\langle  \vec{i},\vec{j} \rangle,\sigma} \left( \hat{c}_{\vec{i},\sigma}^{\dagger}  \hat{c}_{\vec{j},\sigma}^{\phantom\dagger}   + \text{H.c.} \right) }_{\equiv \hat{\mathcal{H}}_t} - \frac{J}{4} 
	\sum_{\vec{i}} \left( \sum_{\sigma} \hat{c}_{\vec{i},\sigma}^{\dagger}  \hat{f}_{\vec{i},\sigma}^{\phantom\dagger}  + 
	                                                        \hat{f}_{\vec{i},\sigma}^{\dagger}  \hat{c}_{\vec{i},\sigma}^{\phantom\dagger}   \right)^{2}   +
        \underbrace{\frac{U}{2}   \sum_{\vec{i}}   \left( \hat{n}^{f}_{\vec{i}} -1 \right)^2}_{\equiv \hat{\mathcal{H}}_U}.
 \end{equation}
This form is included in the general Hamiltonian (\ref{eqn:general_ham})  such that the above Hamiltonian can  be implemented in our program package.  
The  relation to the Kondo lattice model follows  from expanding the square  of the hybridization to obtain: 
 \begin{equation}
 	\hat{\mathcal{H}}  =\hat{\mathcal{H}}_t   
	+ J \sum_{\vec{i}}  \left(  \hat{\vec{S}}^{c}_{\vec{i}} \cdot  \hat{\vec{S}}^{f}_{\vec{i}}    +   \hat{\eta}^{z,c}_{\vec{i}} \cdot  \hat{\eta}^{z,f}_{\vec{i}}  
		-  \hat{\eta}^{x,c}_{\vec{i}} \cdot  \hat{\eta}^{x,f}_{\vec{i}}  -  \hat{\eta}^{y,c}_{\vec{i}} \cdot  \hat{\eta}^{y,f}_{\vec{i}} \right) 
	 + \hat{\mathcal{H}}_U.
 \end{equation}
 where the $\eta$-operators  relate to the spin-operators via a particle-hole transformation in one spin sector: 
 \begin{equation} 
 	\hat{\eta}^{\alpha}_{\vec{i}}  = \hat{P}^{-1}  \hat{S}^{\alpha}_{\vec{i}} \hat{P}  	\; \text{ with }  \;   
	\hat{P}^{-1}  \hat{c}^{\phantom\dagger}_{\vec{i},\uparrow} \hat{P}  =   (-1)^{i_x+i_y} \hat{c}^{\dagger}_{\vec{i},\uparrow}  \; \text{ and }  \;   
	\hat{P}^{-1}  \hat{c}^{\phantom\dagger}_{\vec{i},\downarrow} \hat{P}  = \hat{c}^{\phantom\dagger}_{\vec{i},\downarrow} 
 \end{equation}
 Since the $\hat{\eta}^{f} $- and $ \hat{S}^{f} $-operators  do not alter the  parity [$(-1)^{\hat{n}^{f}_{\vec{i}}}$ ] of the $f$-sites, 
 \begin{equation}
 	\left[  \hat{\mathcal{H}}, \hat{\mathcal{H}}_U \right] = 0.
 \end{equation}
 Thereby,  and for positive values of $U$ ,  doubly occupied  or empty $f$-sites -- corresponding to even parity sites -- are suppressed  by a  Boltzmann factor 
 $e^{-\beta U/2} $ in comparison to odd parity sites.   Choosing $\beta U $ adequately essentially allows to  restrict the Hilbert space to  odd parity $f$-sites.  
 In this Hilbert space $\hat{\eta}^{x,f} = \hat{\eta}^{y,f} =  \hat{\eta}^{z,f} =0$  such that the Hamiltonian (\ref{eqn:ham_kondo}) reduces to the Kondo lattice model. 
%
\subsubsection{$SU(N)$-Hubbard-Heisenberg models}
%
$SU(2N)$-Hubbard-Heisenberg \cite{Assaad04,Lang13} models can be written as:
\begin{equation}
 \hat{\mathcal{H}}  =  
 \underbrace{ - t \sum_{ \langle \vec{i},\vec{j} \rangle }    \left(  \vec{\hat{c}}^{\dagger}_{\vec{i}}  \vec{\hat{c}}^{\phantom{\dagger}}_{\vec{j}} + \text{H.c.} \right) }_{\equiv \hat{\mathcal{H}}_t} \; \; 
\underbrace{ -\frac{J}{2 N}  \sum_{ \langle \vec{i},\vec{j} \rangle  } \left(
           \hat{D}^{\dagger}_{ \vec{i},\vec{j} }\hat{D}^{\phantom\dagger}_{ \vec{i},\vec{j}}  +
            \hat{D}^{\phantom\dagger}_{ \vec{i},\vec{j} } \hat{D}^{\dagger}_{ \vec{i},\vec{j} }  \right) }_{\equiv\hat{\mathcal{H}}_J}
            + 
 \underbrace{\frac{U}{N}  \sum_{\vec{i}} \left(
             \vec{\hat{c}}^{\dagger}_{\vec{i}}  \vec{\hat{c}}^{\phantom\dagger}_{\vec{i}} -  {\frac{N}{2} } \right)^2}_{\equiv \hat{\mathcal{H}}_U}
\end{equation}
Here,
$ \vec{\hat{c}}^{\dagger}_{\vec{i}} =
(\hat{c}^{\dagger}_{\vec{i},1},  \hat{c}^{\dagger}_{\vec{i},2}, \cdots, \hat{c}^{\dagger}_{\vec{i}, N } ) $  is an
$N$-flavored spinor, and $ \hat{D}_{ \vec{i},\vec{j}} = \vec{\hat{c}}^{\dagger}_{\vec{i}}
\vec{\hat{c}}_{\vec{j}}  $.
To use the QMC code of the ALF package  to simulate this model, one will rewrite  the $J$-term as a sum of perfect squares, 
\begin{equation}
        \hat{\mathcal{H}}_J =  -\frac{J}{4 N}  \sum_{  \langle \vec{i}, \vec{j} \rangle }
        \left(\hat{D}^{\dagger}_{  \langle \vec{i}, \vec{j} \rangle  } +  \hat{D}_{  \langle \vec{i}, \vec{j} \rangle }  \right)^2  -
        \left(\hat{D}^{\dagger}_{   \langle \vec{i}, \vec{j} \rangle } -  \hat{D}_{  \langle \vec{i}, \vec{j} \rangle}  \right)^2,
\end{equation}
so to manifestly bring it into the form of the general Hamiltonian(\ref{eqn:general_ham}). 
It is amusing to note that setting the hopping $t=0$,    charge fluctuations  will be suppressed by the  Boltzmann factor $e^{-\beta U /N \left(  \vec{\hat{c}}^{\dagger}_{\vec{i}}  \vec{\hat{c}}^{\phantom\dagger}_{\vec{i}} -  {\frac{N}{2} } \right)^2 } $ 
since in this case  $ \left[   \hat{\mathcal{H}_J}, \hat{\mathcal{H}}_U \right]  = 0 $.
This provides a route to use the auxiliary field QMC algorithm  to simulate -- free of the sign problem -- $SU(2N)$-Heisenberg models in the self-adjoint antisymmetric representation \footnote{ This corresponds to a Young tableau with single column and $N/2$ rows.}.
For odd values of $N$ recent progress  in our understanding of the  origins of the sign problem \cite{Wei16}  allows us to simulate  a set of non-trivial Hamiltonians \cite{Li15,Assaad16},  without encountering the sign problem.
%
\subsection{Performance, memory requirements and parallelization}
\begin{figure}[h]
	\begin{center}
		\includegraphics[scale=.8]{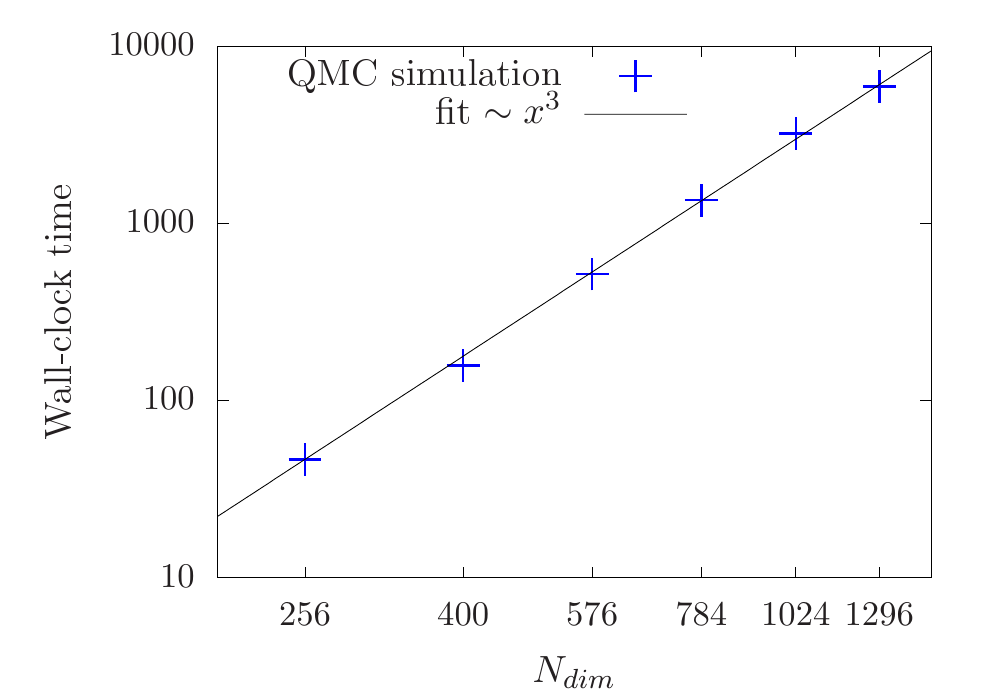}
	\end{center}
	\caption{\label{fig_scaling_size}Volume scaling behavior of the auxiliary field QMC code of the ALF project on SuperMUC (phase 2/Haswell nodes) at the LRZ in Munich. The number of sites $N_{\text{dim}}$ corresponds to the system volume.
	The plot confirms that the leading scaling order is due to matrix multiplications such that the runtime is dominated by calls to ZGEMM. }
\end{figure}

%
As mentioned in the  introduction, the auxiliary field QMC algorithm scales linearly in inverse temperature $\beta$ and cubic in the volume $N_{\text{dim}}$. Using fast updates,  a single spin flip  requires $(N_{\text{dim}})^2$ operations to update the Green function upon acceptance.  As there are $L_{\text{Trotter}}\times N_{\text{dim}}$ spins to be visited, the total computational cost for one sweep is of the order of $\beta (N_{\text{dim}})^3$. This operation  dominates the performance, see Fig.~\ref{fig_scaling_size}. A profiling analysis of our code shows that 80-90\% of the CPU time is spend in ZGEMM calls of the BLAS library provided in the MKL package by Intel. Consequently, the single-core performance is next to optimal.

For the implementation which scales linearly in $\beta$, one has to store $L_{\text{Trotter}}/\texttt{NWrap}$ intermediate propagation matrices of dimension $N\times N$. For large lattices and/or low temperatures this dominates the total memory requirements that can exceed 2~GB memory for a sequential version.

\begin{figure}[h]
	\begin{center}
		\includegraphics[width=0.9\textwidth]{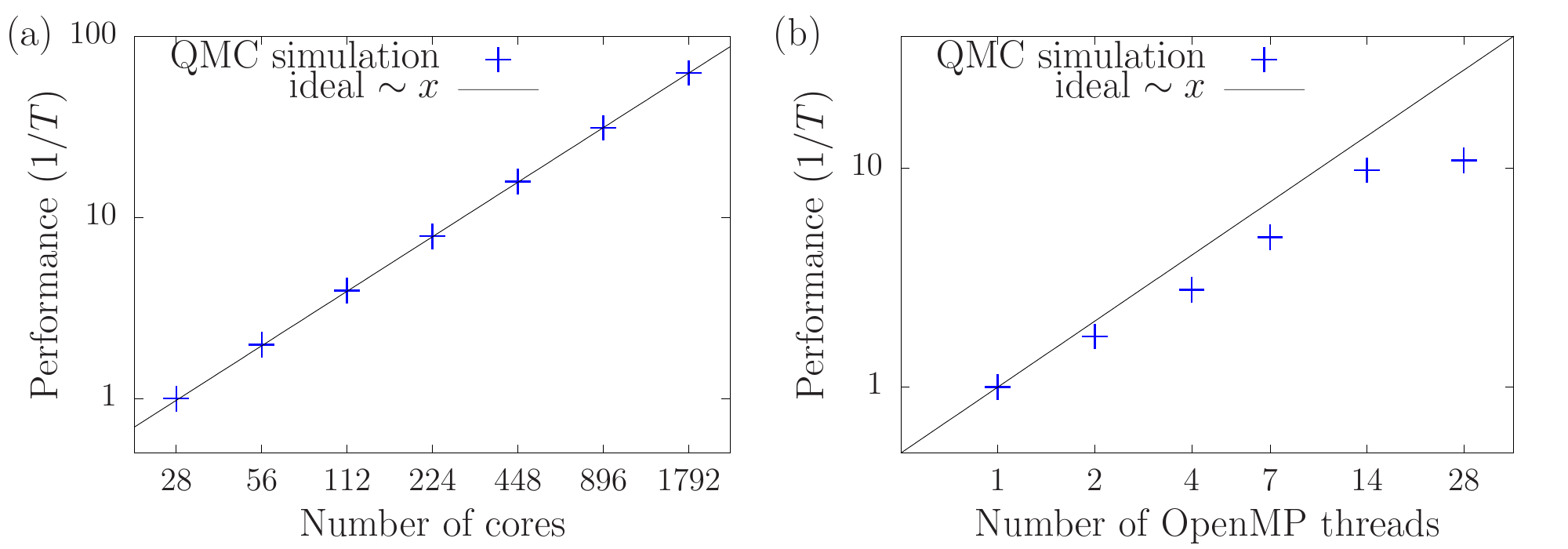}
	\end{center}
	\caption{\label{fig_scaling} MPI (a) and OpenMP (b) scaling behavior of the auxiliary field QMC code of the ALF project on SuperMUC (phase 2/Haswell nodes) at the LRZ in Munich.
		The MPI performance data was normalized to 28 cores and was obtained using a problem size of $N_{\text{dim}}=400$. This is a medium to small system size that is the least favorable in terms of MPI synchronization effects.
		The OpenMP performance data was obtained using a problem size of $N_{\text{dim}}=1296$. Employing 2 and 4 OpenMP threads introduces some synchronization/management overhead such that the per-core performance is slightly reduced, compared to the single thread efficiency. Further increasing the amount of threads to 7 and 14 keeps the efficiency constant. The drop in performance of the 28 thread configuration is due to the architecture as the threads are now spread over both sockets of the node. To obtain the above results, it was crucial to pin the processes in a fashion that keeps the OpenMP threads as compact as possible.}
\end{figure}

At the heart of Monte Carlo schemes lies a random walk through the given configuration space. This is easily parallalized via MPI by associating one random walker to each MPI task. For each task, we start from a random configuration and have to invest the autocorrelation time $T_\mathrm{auto}$ to produce an equilibrated configuration.
Additionally we can also profit from an OpenMP parallelized version of the BLAS/LAPACK library for an additional speedup, which also effects equilibration overhead $N_\text{MPI}\times T_\text{auto} / N_\text{OMP}$, where $N_{\text{MPI}}$ is the number of cores and $N_{\text{OMP}}$ the number of OpenMP threads.
For a given number of independent measurements  $N_\text{meas}$, we  therefore need a wall-clock time given by
\begin{equation}\label{eqn:scaling}
T  =  \frac{T_\text{auto}}{N_\text{OMP}} \left( 1   +    \frac{N_\text{meas}}{N_\text{MPI}}  \right) \,.
\end{equation}
As we typically have $ N_\text{meas}/N_\text{MPI} \gg 1 $, 
the speedup is expected to be almost perfect, in accordance with
the performance test results for the auxiliary field
QMC code  on SuperMUC [see Fig.~\ref{fig_scaling} (a)].

For many problem sizes, 2~GB memory per MPI task (random walker) suffices such that we typically start as many MPI tasks as there are physical cores per node. Due to the large amount of CPU time spent in MKL routines, we do not profit from the hyper-threading option. For large systems, the memory requirement increases and this is tackled by increasing the amount of OpenMP threads to decrease the stress on the memory system and to simultaneously reduce the equilibration overhead [see Fig.~\ref{fig_scaling} (b)]. For the displayed speedup, it was crucial to pin the MPI tasks as well as the OpenMP threads in a pattern which keeps the threads as compact as possible to profit from a shared cache. This also explains the drop in efficiency from 14 to 28 threads where the OpenMP threads are spread over both sockets. 

We store the field configurations of the random walker as checkpoints, such that a long simulation can be easily split into several short simulations. This procedure allows us to take advantage of chained jobs using the dependency chains provided by the batch system.
%
\section{Conclusions and Future Releases}\label{sec:con}
%
In its present form, the  auxiliary field QMC code of the ALF project  allows to simulate a large class of non-trivial models, both efficiently and at minimal  programming cost.  There are many possible extensions which deserve to be considered in future releases.    The model Hamiltonians we have presented so far are imaginary-time independent. This however can be easily generalized  to imaginary-time dependent model Hamiltonians thus allowing, for example, to access  entanglement properties of interacting fermionic systems \cite{Broecker14,Assaad14,Assaad13a,Assaad15}. Generalizations to include global moves are equally desirable. This is a prerequisite to  play with recent ideas of self-learning algorithms  \cite{Xu16a} so as to  possibly avoid the issue of critical slowing down.  Parallel tempering schemes are equally desirable, so as to possibly alleviate   long autocorrelation times.     Most of the above has already been tested  and will  be available in the next  major release of the ALF package.

On the longer term, we foresee further possible developments.  At present, the QMC code of this package is  restricted to discrete HS fields such that  implementations  of  the long-range Coulomb repulsion -- as introduced in \cite{Hohenadler14,Ulybyshev2013,Brower12} -- are not yet included.   Extensions  to  continuous HS fields are certainly possible, but require an efficient upgrading scheme such as hybrid  molecular dynamics \cite{Duane85}.    An  implementation of the  ground state projective QMC method  within ALF is equally desirable.  Efforts in  the above directions will be pursued on the longer term.  

As it stands, programming a new model  certainly requires some  detailed knowledge of the  algorithm. To facilitate access we hope  to maintain an increasing number of  model Hamiltonians in  the ALF  repository.    A further  step is to aim at  cross compatibility with other  major projects, especially the ALPS \cite{ALPS_2.0} project.
%
\section*{Acknowledgments} 
%
We are very grateful to  S.~Beyl, M.~Hohenadler,  F.~Parisen Toldin,  M.~Raczkowski, T.~Sato, J.~Schwab, Z.~Wang, and M.~Weber  for constant support during the development of this project. We equally thank G.~Hager, M.~Wittmann, and G.~Wellein for useful discussions and support.
FFA would also like to thank T.~Lang   and Z.~Y.~Meng for  developments of the auxiliary field code as well as T.~Grover for many discussions. 
\paragraph{Funding information}
MB thanks the Bavarian Competence Network for Technical and Scientific High Performance Computing (KONWIHR) for financial support. FG  and JH thank the SFB-1170 for  financial support under projects Z03 and C01.  FFA thanks the DFG-funded FOR1807 and FOR1346 for partial financial support.
Part of the optimization of the code was carried out during  the  Porting and Tuning Workshop 2016 offered by the Forschungszentrum J\"ulich.
Calculations  to extensively test this package were carried out both on  SuperMUC at the  Leibniz Supercomputing Centre and on  JURECA  \cite{Jureca16} at the J\"ulich Supercomputing Centre.  We thank both institutions for generous allocation of computing time.

\begin{appendix}
\section{License}
%
The  ALF code  is provided as an open source software  such that it is  available  to all and we  hope that  it 
will be useful.  If you benefit from this code  we ask that you acknowledge  the ALF collaboration  as mentioned on our
homepage \url{alf.physik.uni-wuerzburg.de}.   The Git repository at   \url{alf.physik.uni-wuerzburg.de} gives us the tools to 
create a small but vibrant community around the code and provides a suitable entry point for future contributors  and future developments. 
The homepage is also the place where the original source files can be found.
With the coming public release it was necessary to add copyright headers to our source files.
The Creative Commons licenses are a good way to share our documentation and it is also well 
accepted by publishers. Therefore this documentation is licensed to you under a CC-BY-SA license.
This means you can share it and redistribute it as long as you cite the original source and
license your changes under the same license. The details are in the file license.CCBYSA that you should have received with this documentation.
The source code itself is licensed under a GPL license to keep the source as well as any future work in the community.
To express our desire for a proper attribution we decided to make this a visible part of the license.
To that end we have exercised the rights of section 7 of GPL version 3 and have amended
the license terms with an additional paragraph that expresses our wish that if an author has benefitted from this code
that he/she should consider giving back a citation as specified on \url{alf.physik.uni-wuerzburg.de}.
This is not something that is meant to restrict your freedom of use, but something that we strongly expect to be good scientific conduct.
The original GPL license can be found in the file license.GPL and the additional terms can be found in license.additional.
In favour to our users, the ALF code contains part of the lapack implementation version 3.6.1 from \url{http://www.netlib.org/lapack}.
Lapack is licensed under the modified BSD license whose full text can be found in license.BSD.\\
With that being said, we hope that the ALF code will prove to you to be a suitable and high-performance tool that enables
you to perform quantum Monte Carlo studies of solid state models of unprecedented complexity.\\
\\
The ALF project's contributors.
%
\subsection*{COPYRIGHT}
%
Copyright \textcopyright ~2016, 2017, The \textit{ALF} Project.\\
The ALF Project Documentation 
is licensed under a Creative Commons Attribution-ShareAlike 4.0 International License.
You are free to share and benefit from this documentation as long as this license is preserved
and proper attribution to the authors is given. For details see the ALF project
homepage \url{alf.physik.uni-wuerzburg.de} and the file \texttt{license.CCBYSA}.
\end{appendix}



\clearpage

\bibliography{alf-1.0-scipost}

\nolinenumbers

\end{document}